\documentclass{aa}  
\usepackage{graphicx}
\usepackage{xcolor}
\usepackage{txfonts}

\begin{document}

   \title{Long term dynamics around the Didymos-Dimorphos binary asteroid of boulders ejected after the DART impact}
   \titlerunning{Long term dynamics around the Didymos-Dimorphos}
   %\subtitle{subtitle}

   \author{K. Langner
          \inst{1,}\inst{2,}\inst{3}
          \and
          F. Marzari \inst{1,} \inst{2}
          \and
          A. Rossi\inst{1}
          \and
          G. Zanotti \inst{4}
          }

   \institute{
             IFAC-CNR, Via Madonna del Piano 10, 50019 Sesto Fiorentino, Italy  \and 
             Dipartimento di Fisica, Universit\`a di Padova, Italy \and
             Institute Astronomical Observatory, Faculty of Physics, Adam Mickiewicz University, Poznań, Poland
             \and 
             Dipartimento di Scienze e Tecnologie Aerospaziali, Politecnico di Milano, Milano, Italy 
             }

   \date{\today}


\abstract{%In 2022 the DART mission spacecraft impacted the asteroid Dimorphos, the secondary body of the binary Didymos system, to test the possibility of asteroid redirection for planetary defence. The impact ejected a large number of dust particles, rocks and boulders into the space around Dimorphos. In late 2024 European Space Agency will launch the Hera mission towards the Didymos system to further study the outcomes of DART.
In 2022 the DART mission spacecraft impacted the asteroid Dimorphos, the secondary body of the binary Didymos system, ejecting  a large number of dust particles, rocks and boulders. The ESA Hera mission will reach the system in 2026 for post--impact studies and possible detection of orbiting fragments.
}{%To simulate the dynamics of the large boulders ejected by the DART mission over a time span of 4 years to describe the long term-behaviour of objects in the Didymos system and to find possibility if any of those ejected objects remain in the system until Hera mission arrival, due in late 2028.
We investigate the long term dynamics of the large boulders ejected by DART to test if any of these objects survive in orbit until the arrival of the Hera mission. 
}{%A numerical simulation of a large synthetic population of particles was performed using a comprehensive model of the whole binary system including the gravity of non-spherical Didymos and Dimorphos asteroids, Solar gravity and radiation pressure.
To model the dynamics of the boulders we use a numerical model which includes the gravity of non-spherical Didymos and Dimorphos, the solar gravity and the radiation pressure. The SPICE kernels are used to define the correct reference frame for the integrations. 
}{
The dynamics of the boulders is highly chaotic and 1\% of the initial boulders survive at least for 4 years on quasi--stable orbits. These orbits are characterised by wide oscillations in eccentricity in antiphase with those in inclination (including spin flips), a mechanism similar to the Kozai one. This behaviour may protect these bodies from close encounters with both asteroids. We also compute the distribution on the surfaces of the asteroids of sesquinary impacts which may influence the dust emission, after the initial DART impact, and the surface composition of the asteroids. 
%A small fraction of the simulated objects survived in the system until the end of the simulation. Other objects either re-impact with one of the asteroids or escape the system. The early dynamical evolution of the boulders is very chaotic and small changes in the initial conditions can dramatically modify the results for the specific test particle. If an object survives long enough its orbit can stabilise but its dynamics is complicated and can include effects similar to the Kozai mechanism, with high increases of the orbital eccentricity and inclination and also orbital orientation flips.
%The re-impact events occur during the whole time span of the simulation. The locations of the re-impacts have non uniform distribution over the surface of the asteroids and their typical velocity is around 20 cm/s (with maximum up to 80 cm/s) for Didymos and 10cm/s for Dimorphos. 
}{%We found that the probability for large boulders to remain in the system until the Hera arrival is low, but greater than zero. Most of the objects are removed in the early phases of the evolution. An increase in escapes from the system when Didymos is close to its perihelion is noticed too.
The probability of observing boulders by the mission Hera is small but not negligible and an almost constant flux of escaping boulders is expected in the coming years since their lifetime after the DART impact covers a large time interval. Most of re--impacts on Dimorphos occur in the hemisphere opposite to the impact site, preferentially close to the equatorial plane. }

\maketitle
\section{Introduction}
On September 26, 2022 the DART probe impacted Dimorphos, the small moon of the binary 65803 Didymos asteroid system \citep{NatureDaly, NatureCheng}. The event generated a large ejecta plume that was imaged first be the small LICIACube cubesat \citep{DOTTO2021105185,CAPANNOLO2021208}, released by DART 15 days before the impact, during its fly-by. The images revealed a complex plume with a cone-like structure, crammed with dust, clumps of objects, filaments and larger boulders \citep{NatureDotto,Desh}. The event and the resulting ejecta were also observed by several space and ground based observatories which allowed the characterisation of the plume evolution and, in particular, the formation of a long tail of debris, pushed by the solar radiation pressure, stretching for thousands of km \citep{NatureLi}. Further observations by the Hubble Space Telescope revealed the presence of a large population of dozens of boulders moving in the surroundings of the system \citep{Jewitt}. Assuming a geometric albedo of 0.15, these large objects span a range of dimensions between about 4 to $\approx 7$ m in diameter. Jewitt and co-authors estimated sky-plane velocities in the range between 10 cm/s to more than 67 cm/s, without a significant correlation between the size and the velocity of the boulders. Since the escape velocity from the binary system is about 24 cm/s, these velocities are consistent with objects that do not re-impact Dimorphos right after the ejection and that can stay within the system for a comparatively long time span on highly perturbed chaotic orbits. 

In \citep{Rossi} a theoretical analysis of the behaviour of these long survivors was performed highlighting the complex dynamical environment and the possible fate of the ejected large particles (of the order of tens of cm). In this work we concentrate on a synthetic population of objects similar in size to the observed boulders. By means of a comprehensive model, described in Sect.~\ref{sec:DynMod}, we explore the long term dynamics of this population with the aim of characterising, also from the statistical point of view, its behaviour with respect to the possible end-states, including re-impact against the two asteroids or escape from the system. 

Within the international collaboration for planetary defence, the Hera probe, from the European Space Agency, will be launched in October 2024 and will reach the Didymos system in late 2026 \citep{Michel}. Hera will rendezvous the system and will perform a detailed in-situ analysis of the asteroids to fully characterise the outcomes of the DART impact. The possibility that some of the ejected boulders could survive up to the arrival of Hera poses some caveats both on the safety of the probe and on the best way to possibly observe them. In Sect.~\ref{sec:Discussion} some considerations on this subject are presented.

\section{Dynamical Model}\label{sec:DynMod}

To investigate the long term dynamics of the ejecta fragments, once they leave the surface of Dimorphos, we have numerically integrated their trajectories. During their evolution, the fragments may have frequent close encounters with either components of the binary asteroid, making their orbits highly chaotic. The 15th-order Radau numerical integrator \citep{radau1985}, which has a variable step size and is very accurate in modelling close approaches, is very well suited to handle this problem.

The gravitational attraction of both Didymos and Dimorphos is computed taking into account their non-spherical shape. In the model two computational methods are considered: a less CPU time-consuming analytical approach based on the MacCullagh formula \citep{murray2000} or a more cumbersome polyhedral approach \citep{werner1994,werner_scheeres1997}. In short, the MacCullagh formula approximates the gravitational potential of an irregular body as

\begin{equation}
V= - \frac{GM}{r} -
 \frac{G(A+B+C-3I)}{2 r^3}
\end{equation}

where $A,B,C$ are the inertia moments of the body along the principal axes,  $x,y,z$ are coordinates measured along those axes, $r$ is the distance from the center of a body and 
\begin{equation}
I = \frac{(Ax^2+By^2+Cz^2)} {r^2}
\end{equation}

For a triaxial ellipsoid, the inertia moments are related to the principal semi-axes $a,b,c$ through the equations: 

\begin{eqnarray*}
%\begin{split}
A & = & \frac{4}{15} \pi \rho abc (b^2 + c^2) \\
B & = & \frac{4}{15}\pi \rho abc (a^2 + c^2) \\
C & = & \frac{4}{15}\pi \rho abc (a^2 + b^2)
%\end{split}
\label{Maccullagh3}
\end{eqnarray*}

Even if this algorithm is slightly inaccurate close to the body surface, where the polyhedron model would be more precise even if significantly slower, we have adopted it because of the strong chaotic nature of the trajectories. The values of $a,b,c$ for both Didymos and Dimorphos are derived from the shape of the objects as observed by DART \citep{NatureDaly} and obtained from the SPICE kernels. We recall that the "SPICE" (Spacecraft, Planet, Instrument, C-matrix, Events), maintained by the NASA's Navigation and Ancillary Information Facility (NAIF), is an observation geometry information system designed to assist scientists in planning and interpreting scientific observations from space-based instruments aboard robotic planetary spacecraft (https://naif.jpl.nasa.gov/naif/data.html). As an example, typically the SPICE provide positions and velocities of
planets, satellites, comets, asteroids and spacecraft in several different reference systems.

The numerical integration of the fragments' orbits is performed in a body-fixed reference frame centred on Dimorphos, rotating with the asteroid's angular velocity. This reference system is coded within the SPICE kernels as \texttt{IAU\_DIMORPHOS}. Therefore the apparent forces due to the frame rotation are included in the force equation giving the acceleration: 

\begin{equation}
\mathbf a_{app}=-\boldsymbol\omega\times(\boldsymbol\omega\times \mathbf r)- 2\boldsymbol\omega \times \mathbf v,  \\
\end{equation}
where $\boldsymbol\omega$ is the rotation vector of the reference frame and $\mathbf r$ and $\mathbf v$ are the position and velocity vectors in this rotating frame.

The solar tide force is computed in the model as difference between the Sun gravitational force acting on the fragment and that on Dimorphos. The radial vectors with respect to the Sun are derived from the heliocentric orbit of the binary asteroid, once again as provided by the SPICE kernels in the appropriate integration reference system. 

The solar radiation pressure is also included in the trajectory computation with the standard formulation 

\begin{equation}
\mathbf{F} = \frac{S A}{c } Q_{PR} \mathbf{s}
\end{equation}

where $S$  is the solar radiation flux density at the heliocentric distance of the body, $A$ is the fragment geometrical cross section, and $Q_{PR}$ is a dimensionless coefficient determining the amount of radiation that is either reflected, absorbed and re-emitted \citep{burns1979}. The versor $\mathbf{s}$ is directed in the anti-solar direction. 
 
The fragments can be modelled as ellipsoidal rotating particles, therefore posing a time-varying 
cross section for the computation of the solar radiation pressure (see \cite{Rossi} for details). We note that in the current study, focused on large boulders with a low area-to-mass ratio, the effect of the particles shape is not relevant and therefore spherical particles were considered. The algorithm allows also to treat collisions between the fragments. These collisions are mostly effective for small objects in the first phases of the plume evolution (see e.g. \cite{FanPSJ}), therefore, since we are interested in the dynamics of boulders following the first stormy post-collision moments, we assume they are not significantly perturbed by impacts against the smaller fragments and do not consider this effect here.

The accurate heliocentric orbit of the Didymos system and the binary orbital dynamics is based on the JPL Ephemeris, as derived from the post DART impact SPICE kernels.

\section{Long Term Dynamical Behaviour of the Boulders}

\subsection{Initial conditions}

In trying to understand the general behaviour of the ejected boulders we have performed a simulation including 10000 synthetic large objects over an interval of four years after the DART impact, to cover the Hera arrival time--span. The diameters of the simulated objects range from  4 to 10 m and their number, for each size bin in which the size range has been divided, is derived by using a power law distribution with slope $q=-3.9$. This distribution is derived from the  Hubble images taken by \cite{Jewitt}. The boulders are all started near the location of the DART impact site (\cite{NatureDaly}, retrieved within the model from the SPICE kernels) and a small delay in their release from the crater is assumed (\cite{raducan23_inpress}), within an interval of time of about 10 minutes after the impact. The initial velocities with respect to the surface of Dimorphos are encompassed between 6 and 13 cm/s with a uniform random distribution. The directions of the velocity vectors is defined by the ejection angle, measured from the impact site plane. Such angle spans the range between 72$^{\circ}$ and 44$^{\circ}$ \citep{NatureLi,NatureDotto,Desh}, assuming a linear distribution from the innermost ejecta with the steeper angles to the outermost ones, including also the effects of an oblique impact \citep{zanotti2020science,cintala1999,anderson2003}, based on values estimated from numerical experiments \citep{raducan23_inpress}.
The narrow range of velocities and the large size adopted for the bodies were chosen with the intent of maximising  the number of particles potentially surviving for a long time. Note that the number of boulders we simulate is at least two orders of magnitude larger than the expected number. 
The reason of this choice is to have a significant statistics of bodies which are trapped in the gravity field of the binary asteroid for a long time. 

\subsection{General statistics}
The boulders ejected after DART are divided into three groups depending on their final state in the simulation. The first group includes the objects which escape from the system ending in a heliocentric orbit. To determine if an object is outside the Didymos-Dimorphos gravitational influence we took a simple approach based on the classical definition of a Hill radius:
\begin{equation}
    r_H=a_D \left(1-e_D\right) \left(\frac{1}{3 M_\odot}\right)^\frac{1}{3}\approx 68km,
\end{equation}
where $a_D$ is the heliocentric Didymos semi-major axis, $e_D$ is its eccentricity and $M_\odot$ is the mass of the Sun. Because of the high eccentricity of Didymos we found $1 r_H$ too restrictive, so to add extra margin we decided to define the escaping object as and object that is at a distance greater than $5 r_H$ from the  Didymos-Dimorphos barycenter. When a particle reaches that distance, it is considered definitively escaped and removed from the simulation.
This is an empirical criterion adopted to define a large enough distance beyond which we can safely claim that the body is not gravitationally tied to the two asteroids. For this reason, even if the Hill's sphere changes with the heliocentric distance, this does not affect in any way our results.
Boulders colliding either with Didymos or Dimorphos belong to group two. Numerically, an object collides with one of the two asteroids if it enters the ellipsoid representing the surface of the asteroid. Finally, group three includes all objects that survive until the end of the simulation i.e. objects that do not escape nor impact the asteroids during the four-year time span of the simulation.

\begin{figure*}
\centering
\includegraphics[width=\textwidth]{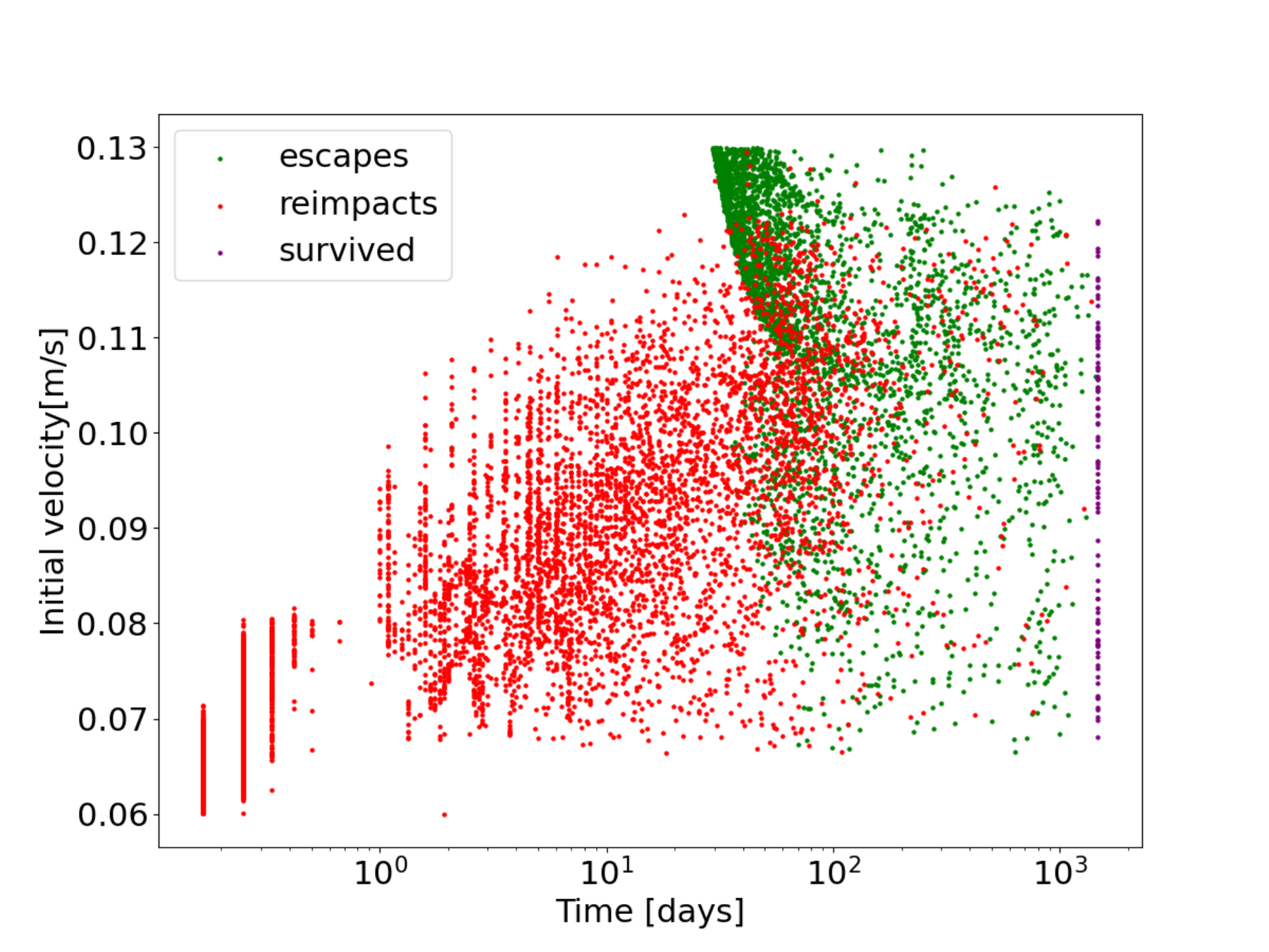}
\caption{The status of all simulated objects. The time shows the moment when the object left the simulation by re-impacting or escaping.}
          \label{fig:stat}%
\end{figure*}

Figure~\ref{fig:stat} illustrates the boulders lifetime vs the initial ejection velocity from Dimorphos' surface. The red dots show the fragments belonging to group two which re-impact the asteroid after some time, the green dots represent the bodies which escape from the system (group one) while the purple points are the boulders surviving until the end of the simulation (4 years) therefore belonging to group three. According to Fig.~\ref{fig:stat} most of the low velocity boulders re-impact Dimorphos on a short timescale while only few escape. On the other hand, the higher velocity fragments are more prompt to escape after timescales longer than about 100 days and a few either impact Dimorphos or, preferentially,  Didymos. The long surviving bodies which accumulate in the rightmost part of the plot, do not show a significant dependence of the ejection velocity.
The statistics of the final state of the boulders is shown in Table~\ref{tab:objects}. It is worth noticing that the percentages of objects colliding and escaping changes over time, where the re-impacts with Dimorphos dominate the early stages of the simulation, while escapes and impacts with Didymos become more frequent later. Most of the particles are removed during the first 100 days and only 1565 of the original 10000 remained. At the end of the simulation, after 4 years, 92 boulders survived which is slightly less than $1\%$ of the initial objects. In the next section, we describe the evolution in each of the groups of boulders in more detail.

\begin{table}
  \caption[]{The state of the particles in simulation.}
     \label{tab:objects}

     \begin{tabular}{lccc}
        \hline
        \noalign{\smallskip}
        Status      &  Total &$t<100d$ &$t>100d$ \\
        \noalign{\smallskip}
        \hline
        \noalign{\smallskip}
        Impact Didymos& 1863 &759&225    \\
        Impact Dimorphos& 5429&5285&144     \\
        Escape& 3417&2313&1104    \\
        Survived&92&1565&92 \\
       \hline
     \end{tabular}

\end{table}

\subsection{Escaping boulders}

The images in \cite{Jewitt} show a population of boulders left within the Didymos system after the DART impact. 
As mentioned before, in our simulation we restricted the initial velocities to a narrow range which is slightly lower than escape velocity from Dimorphos up to slightly greater than escape velocity from Didymos-Dimorphos system. Most of our bodies have velocities between this two limits, since we are interested in the long surviving bodies. Therefore our escaping population might follow a different dynamical path, possibly dominated by close encounters with the two asteroids, with respect to the some of the population observed by Hubble which partly consists of boulders ejected with a higher initial velocity.

\begin{figure}
   \centering
   \includegraphics[width=\hsize]{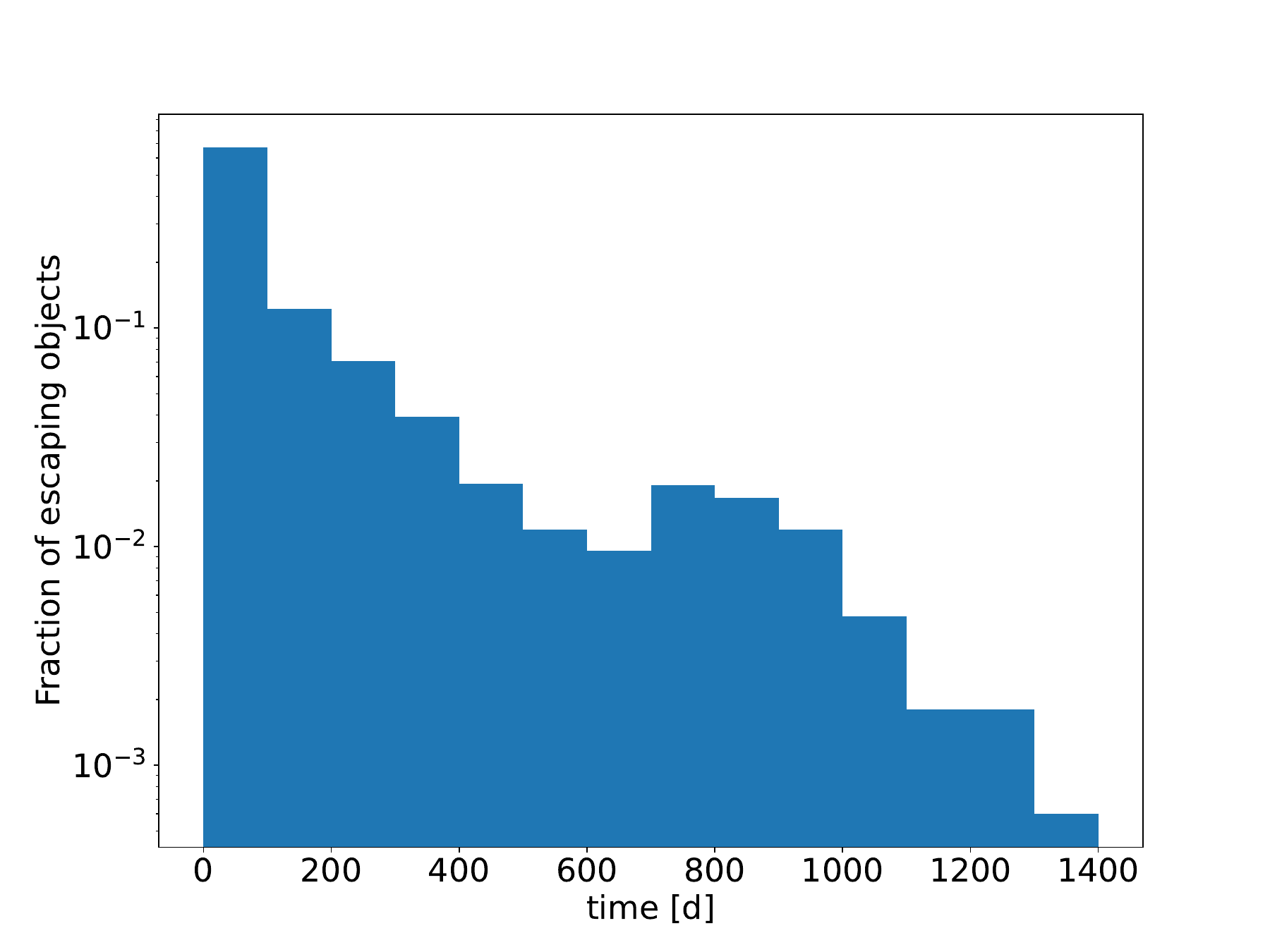}
   \includegraphics[width=\hsize]{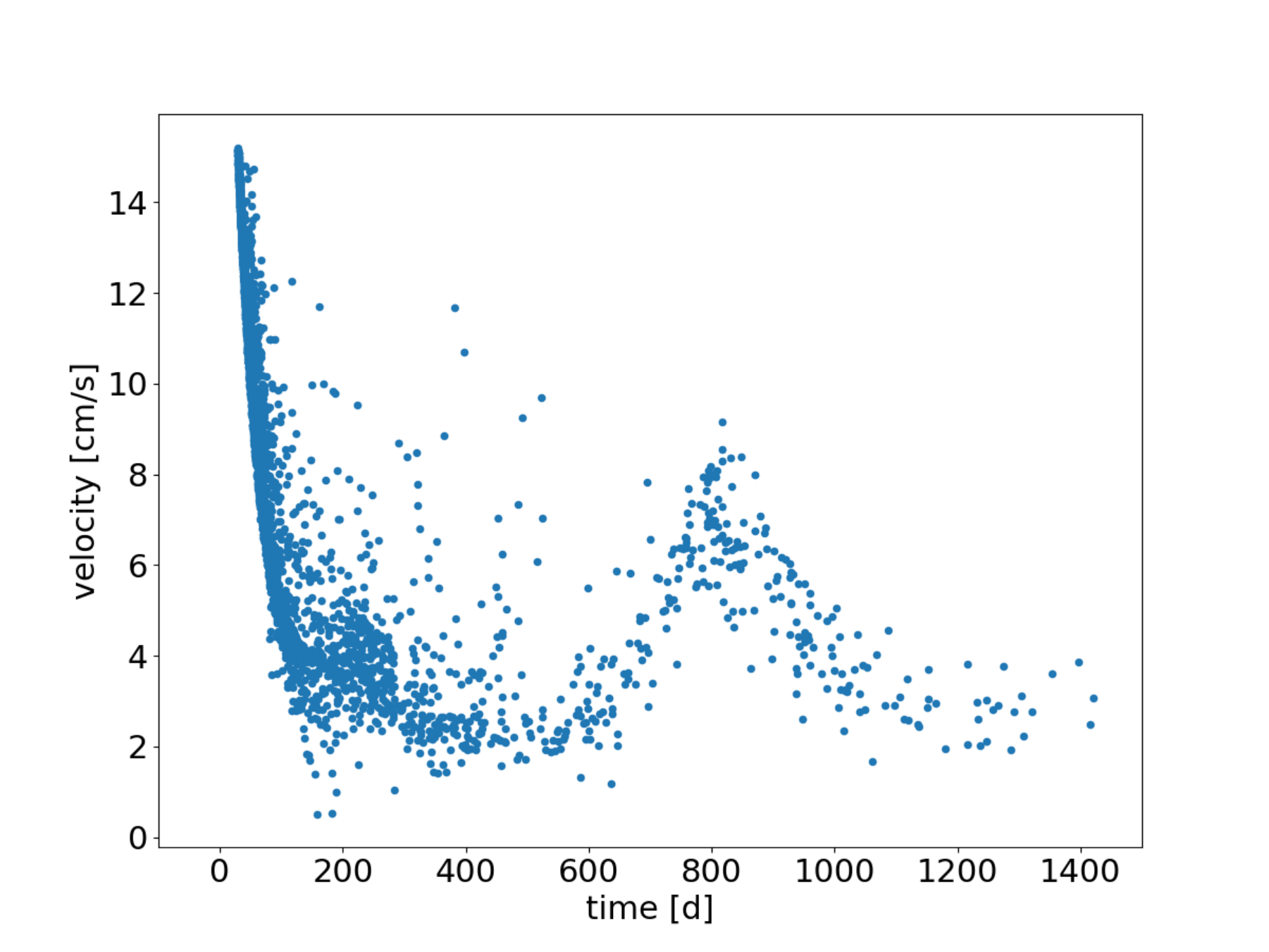}
      \caption{Objects escaping from Didymos system. The top plot shows the total number of escaping boulders in the simulation (normalised, where 1 is the total number of escapes). In the bottom plot each point represents a single boulder showing its escape time and velocity relative to Didymos barycenter at the escape time.
              }
         \label{fig:escape_number}
\end{figure}

If we look at the total number of escaping particles (Fig.~\ref{fig:escape_number}), we can see the peak in the first 100 days of simulation, then the number of escape events decreases. The initial peak consists in both objects that initially have high enough velocity to escape the system and the slower ones that achieved the required velocity due to gravitational interaction with asteroids (see Fig.~\ref{fig:stat}). The latter escapes are a product of the dynamical evolution of boulders in the Didymos-Dimorphos system. The number of escape events decreases exponentially as a result of the decrease in the total number of remaining particles. 
In the simulation, we notice an increase in the number of escapes and in the velocity of escaping particles with a local maximum around 800 days after the Dart impact. This coincides with the moment when Didymos is in its perihelion. The lower Sun-Didymos distance causes stronger Solar tides, and results in boulders on weakly bounded orbits to escape more likely and with higher relative velocity.  We note that this result might suggest to have a dedicated observational campaign at the time of the next perihelion passage to check if some escaping boulders could be detected.

\subsection{Re-impacting boulders}

The investigation of re-impacting boulders on the surface of either Didymos or Dimorphos is relevant under many aspects. They may perform some orbits in the system before colliding and, therefore, they can be classified as sesquinary. It has been shown that this type of impacts may significantly affect the surface of Phobos producing low velocity crater chains \citep{nayak2016}. 
In the case of boulders re--impacting the surface of Didymos/Dimorphos, the collision velocity is low, of the order of  tens of cm per second. It is possibly too low to produce significant secondary craters on the surface of the bodies. However, the sesquinary impacts can lift dust from the surface of Dimorphos and Didymos which will be subsequently removed by solar radiation pressure. This second generation dust may explain the extended period of time during which a tail behind Didymos has been observed (see also \cite{Moreno2023,Ferrari}). They may also change the surface composition by exposing, after the dust lift, deeper portion of the surface. These boulders would also increase the overall boulder counting in the densely re-impacted regions, giving an additional contribution to the surface composition alteration. These boulders are in fact produced by the DART impact which excavated and partly melted material from deeper regions of the asteroid body and when they land on the surface of the asteroids they bring material with a composition which differs from that of the surroundings. Finally, re-impacts can be relevant for the evolution of the mutual orbit of the binary components (e.g., see \cite{Rich2022}). For this reason it is important not only to estimate the frequency of this events but also to find the regions on the surface of both asteroids where most of the primary fragments impact.

\begin{figure}
   \centering
   \includegraphics[width=\hsize]{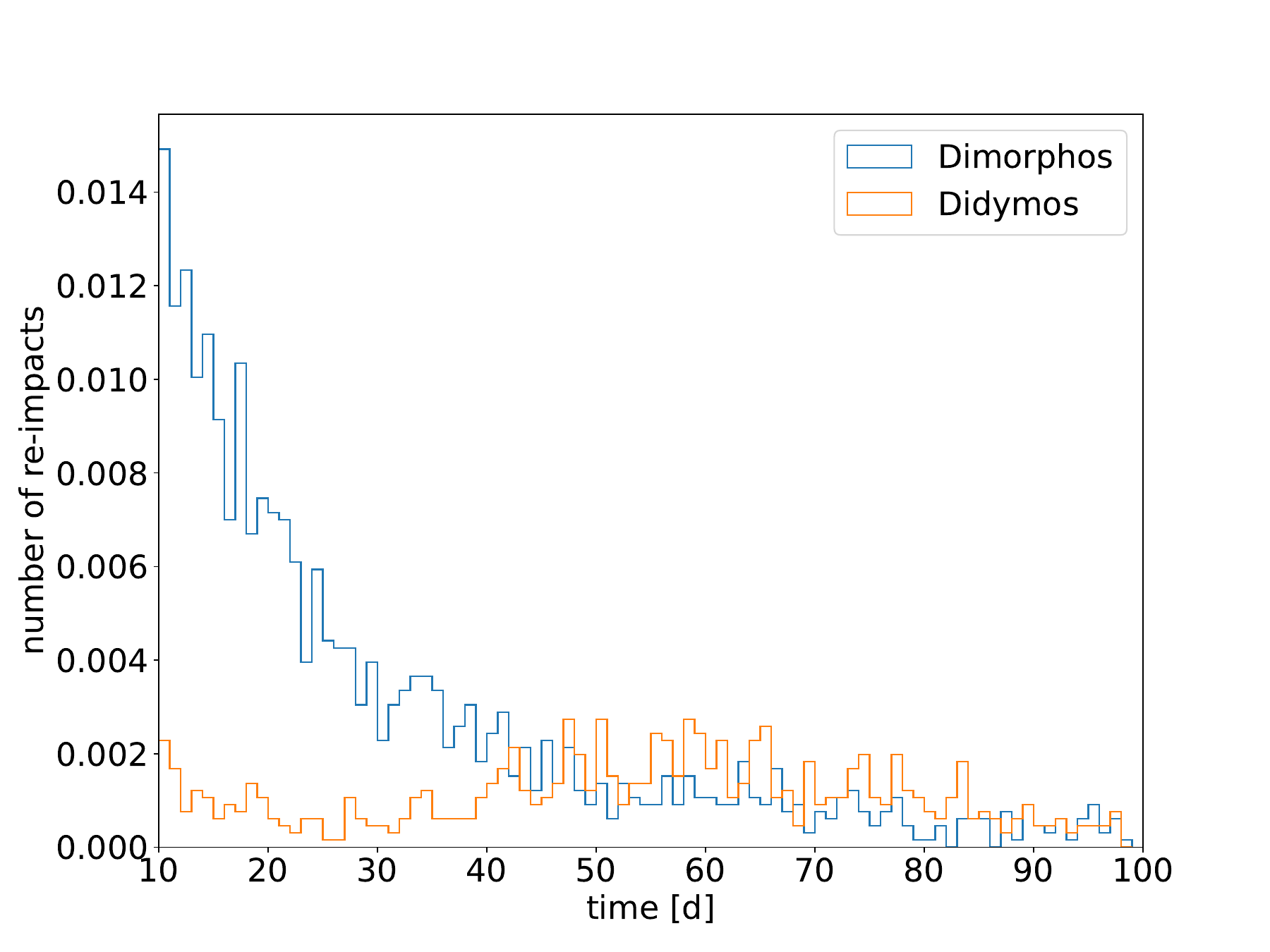}
   \includegraphics[width=\hsize]{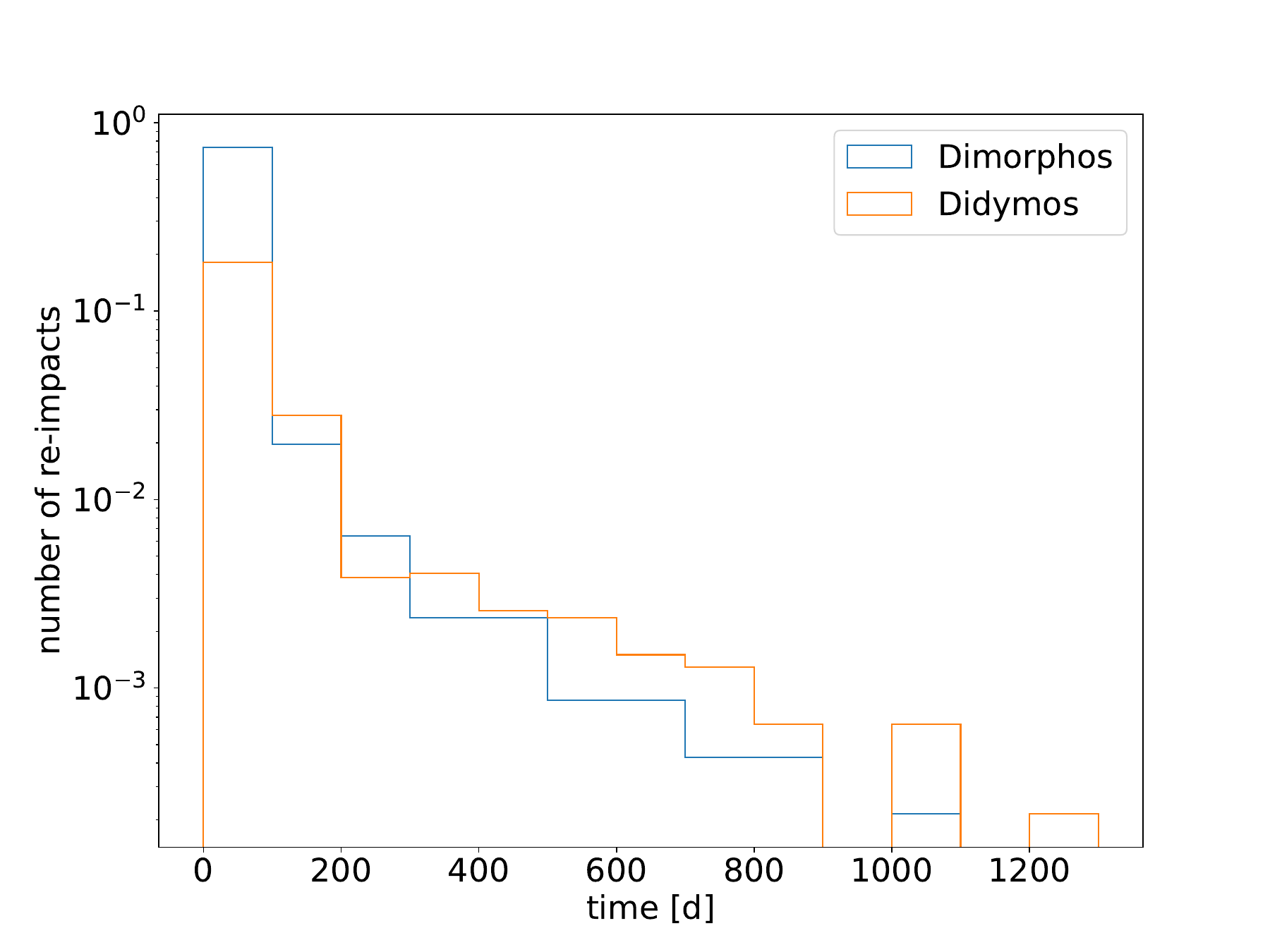}
      \caption{Bottom panel: total number of re-impact events in the whole simulation time-span (logarithmic scale on the y-axis). Top panel: detail of the period from 10 to 100 days after the DART impact, using a linear scale on the y-axis. The blue line shows the number of re-impacts with Dimorphos while the orange line represents the re-impacts with Didymos. Both plots are normalised such as 1 is the total number of re-impacts on both asteroids.
              }
         \label{fig:reimpact_number}
\end{figure}

A first group of re-impacting particles consists of those which are too slow to escape the gravity of Dimorphos and collide with it after a few hours from the DART impact. They re-impact on Dimorphos with a velocity close to their ejection speed. A second group of bodies leave Dimorphos and orbit Didymos for a while before colliding with either of the two asteroids later in the simulation. As shown in Fig.~\ref{fig:reimpact_number} in the first 40 days most of the boulders impact with Dimorphos while later on the impacts with Didymos become more likely. 

\begin{figure}
   \centering
   \includegraphics[width=\hsize]{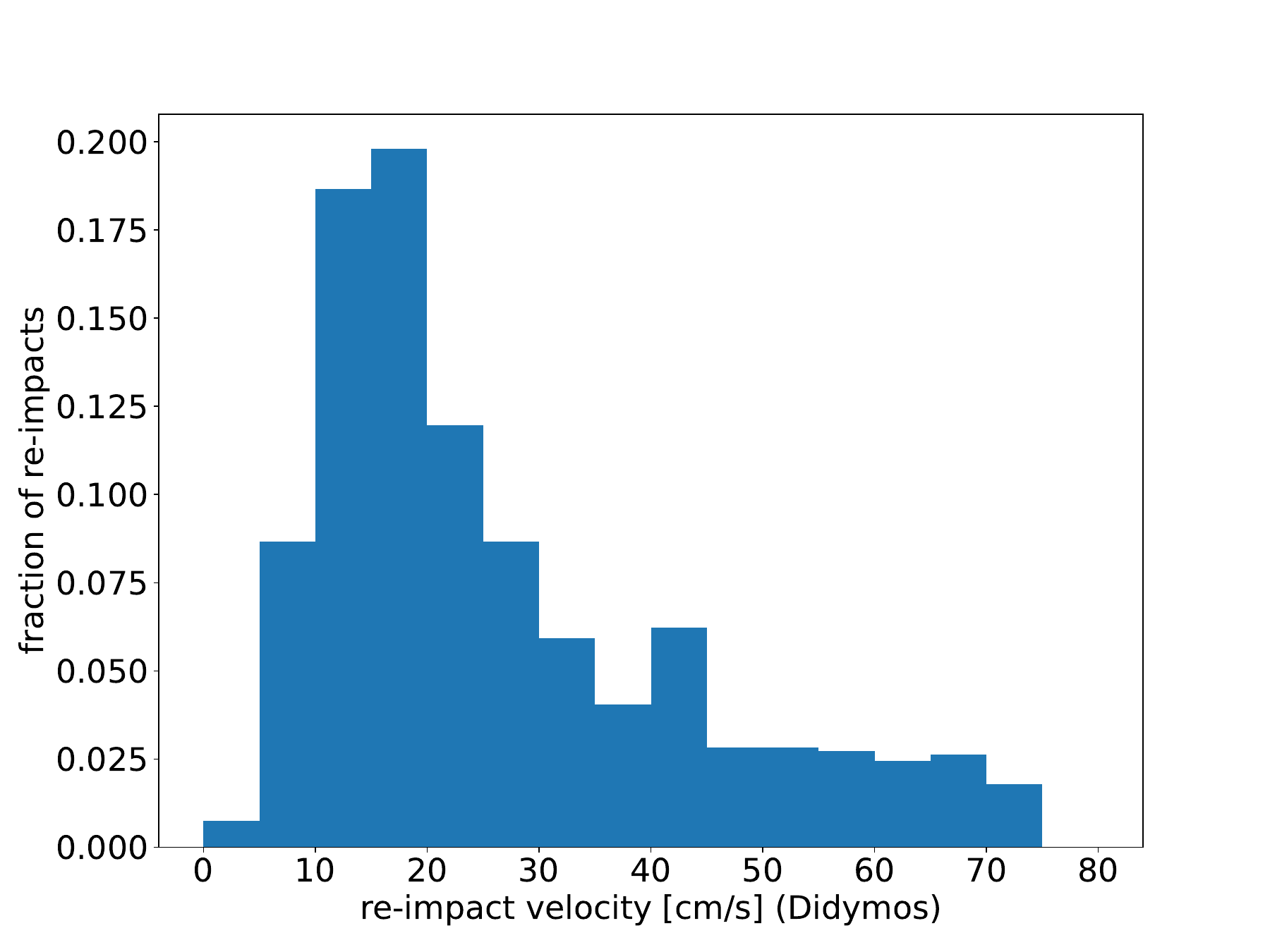}
   \includegraphics[width=\hsize]{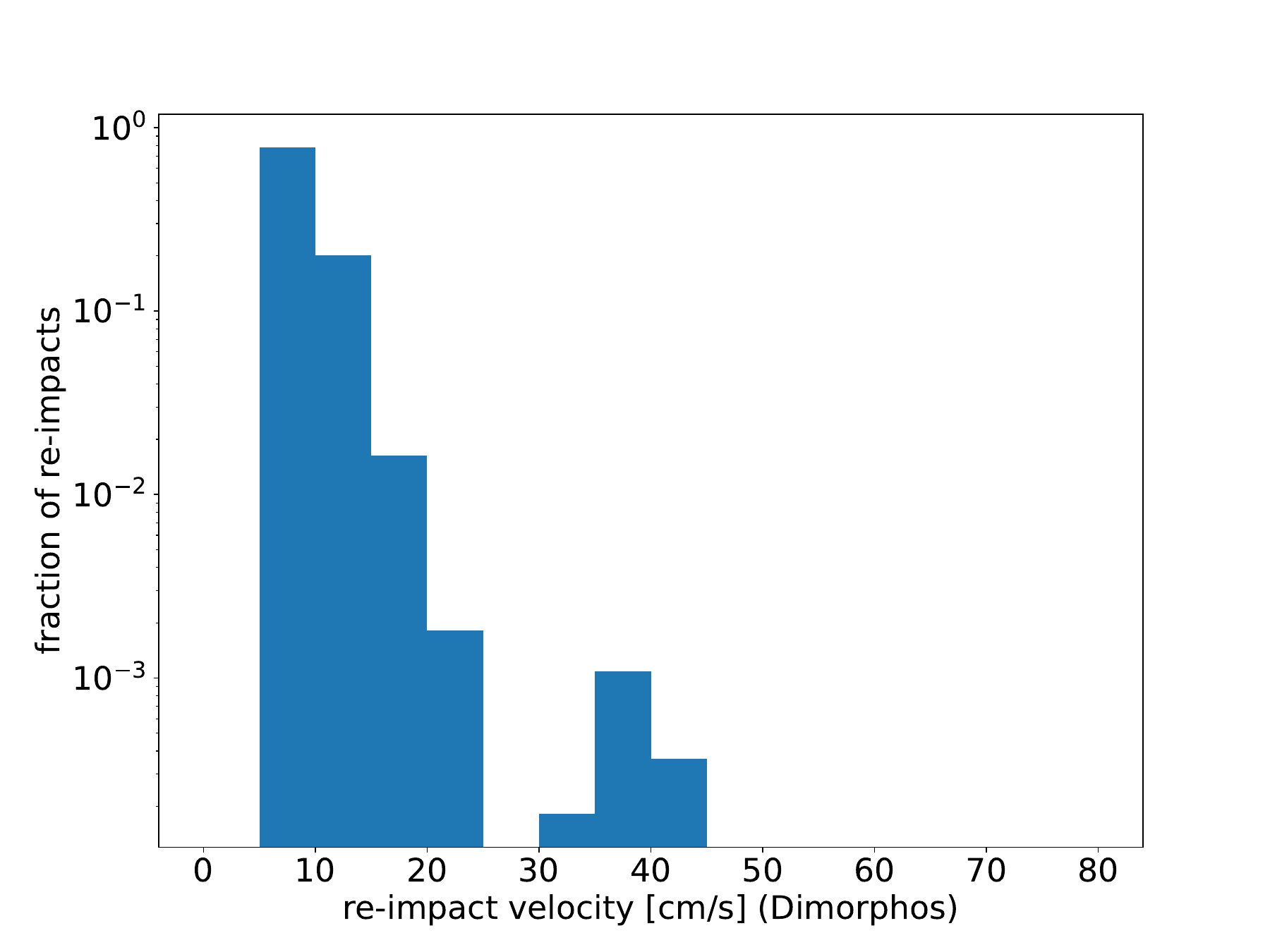}
      \caption{The velocity of boulders re-impacting with Didymos (top) and Dimorphos (bottom). For Dimorphos the logarithmic scale is used for better readability.
              }
         \label{fig:reimpact_velocity}
\end{figure}

For each of the impacts recorded in the simulation we calculate the velocity (Fig.~\ref{fig:reimpact_velocity}). The collisions with Dimorphos are generally slower with mean velocity equal to 8.9 cm/s. Most of the slowest impacts occur early and if we consider only impacts after 10 days the mean impact velocity increases to 10.9 cm/s. The second peak in the histogram of the collision speed on Dimorphos is due to boulders which were injected in a prograde orbit around Didymos and re-impact Dimorphos at a later time, with the earliest one registered 65 days after the beginning of the simulation. This second smaller group of fragments collides with the asteroid with an average higher velocity, about 40 cm/s. 

The population of boulders colliding with Didymos has a higher mean velocity, about 26.6 cm/s, due to the stronger gravity of Didymos. In contrast to Dimorphos the re-impact velocity is not correlated with the orbit being prograde or retrograde.

\begin{figure}
   \centering
   \includegraphics[width=\hsize]{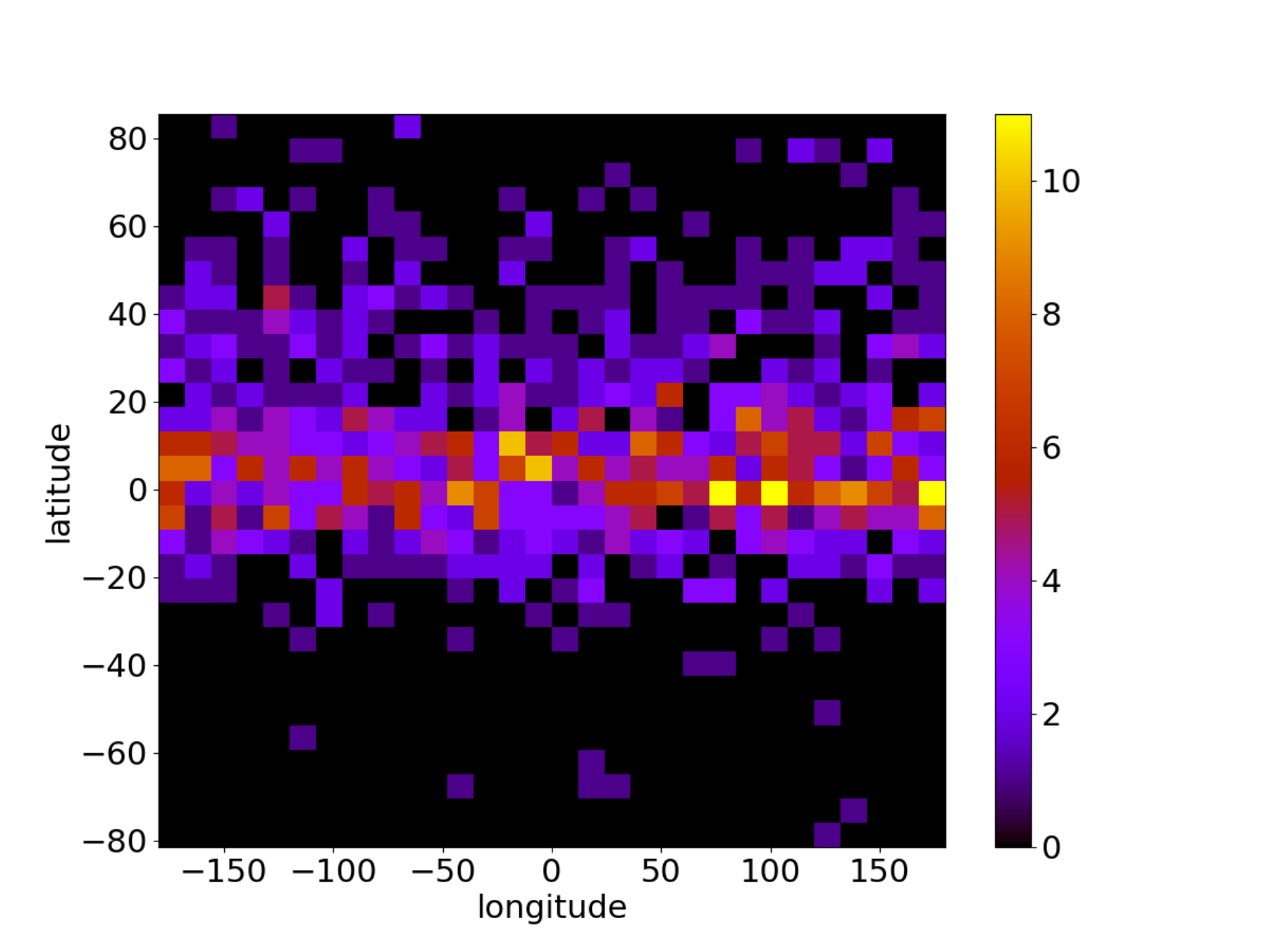}
   \includegraphics[width=\hsize]{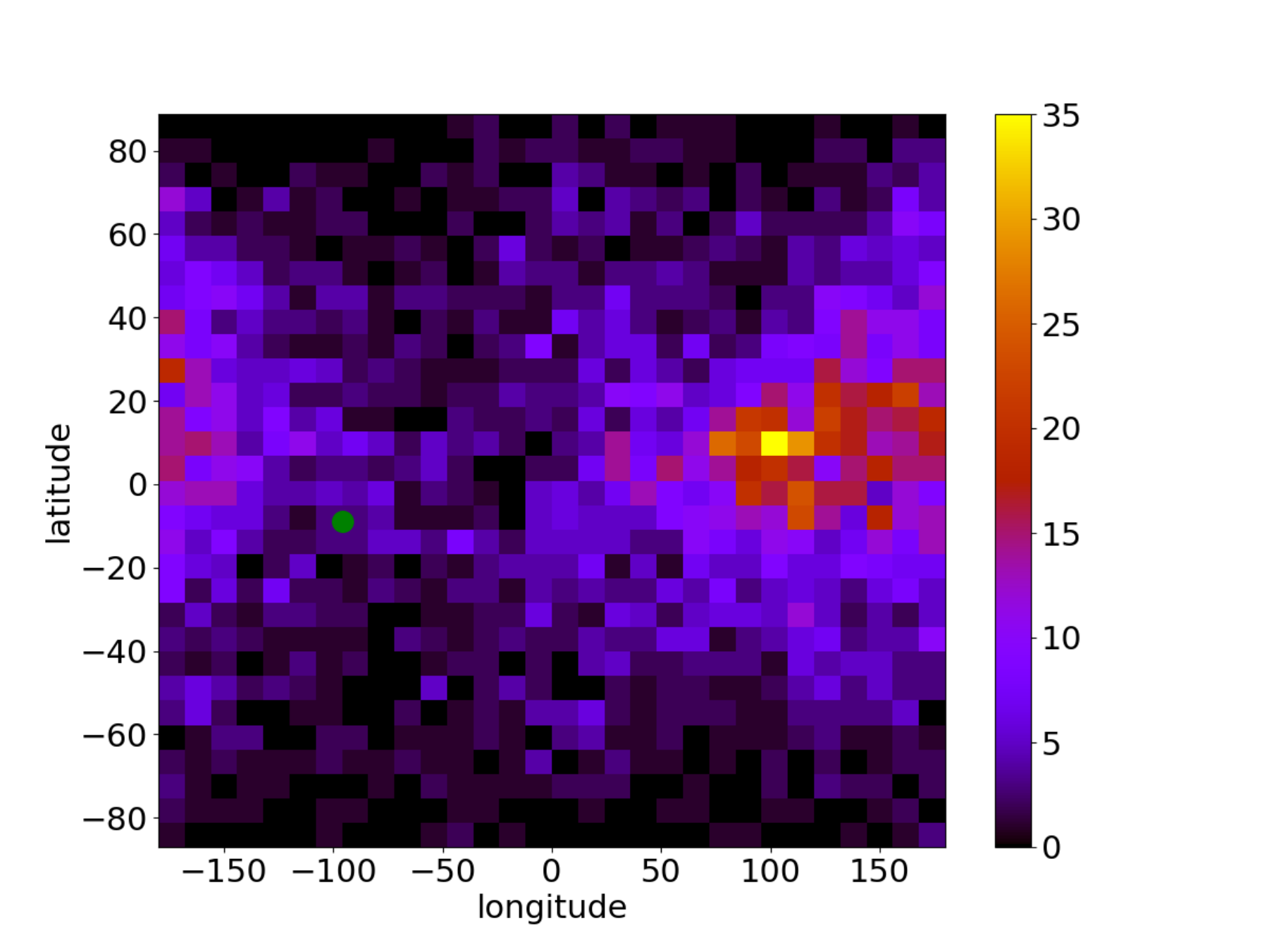}
      \caption{Location of re--impacts on Didymos (top) and on Dimorphos (bottom). Only the re-impacts later than 24 hours after DART impact are included. On the bottom the green dot marks the  DART impact location and the ($0^{\circ}$,$0^{\circ}$) point is the direction towards Didymos and north pole ($+90^{\circ}$ latitude) is in the direction of the rotation axis, while the ($-90^{\circ}$,$0^{\circ}$) is the direction of Dimorphos orbital motion. Most of the reimpacts concentrates near the point ($90^{\circ}$,$0^{\circ}$) which is opposite to the direction of orbital motion of Dimorphos and close to opposite to the DART impact location.}
         \label{fig:reimpact_location}
\end{figure}

The statistical distribution of impacts on the surface of either asteroids is shown in Fig.~\ref{fig:reimpact_location}. To define the latitude and longitude we assumed an ellipsoidal shape of both asteroids and used positive rotation axis direction as the North pole (note that both rotations are retrograde with respect to the ecliptic plane or Didymos heliocentric orbital plane). The longitude angle increases in the direction of rotation, and for Dimorphos the prime meridian is directed towards Didymos. On Didymos the impact locations are uniformly scattered around the equator with a slightly higher probability of impacting on the northern hemisphere. The equatorial preference is related to the initial orbits of most boulders which share an inclination similar to that of Dimorphos which is close to the equatorial plane of  Didymos. 
A small number of bodies re-impact Didymos at later times during which their orbits had time to evolve in inclination 
due to the perturbation in the system. 

To compute the distribution of re-impacts on the surface of Dimorphos we assumed that its rotation is tidally locked to its orbital motion. However, it is possible that after DART Dimorphos entered a state of tumbling (e.g. \cite{Agrusa2021,Meyer2023}). However, since most of the re-impacts occur on a very short timescale after the DART collision, our impact map is still roughly reliable even in the case of the switch to tumbling rotation. Note that, even in the case of a principal axis rotation, Dimorphos could still experience a libration with an amplitude on the order of tens of degrees \citep{Agrusa2021,Meyer2023} that might slightly alter the simulated distribution on our impact map. The re-impact distribution on Dimorphos significantly differ from that on Didymos as expected since most of re-impacting bodies do not come from fragments that have experienced a significant orbital evolution, such as in the case of Didymos. The latitude of the impacts is no longer limited to near equatorial regions and the distribution of re-impact longitudes becomes non-uniform with a strong  preference for positive values.
%(\textcolor{red}{FRANCESCO: a better explanation is needed, in particular if the longitudes are computed respect the Didymos Dimorphos line, positive values mean that they re-impact not from behind........not clear KL: added more explanation in text and image caption describing the used lat long angles. I used the x,y,z definitions of the IAU-DIMORPHOS frame and calculated the spherical angles.})
The re-impacts mostly occur on the opposite hemisphere with respect to the DART impact site. To explain this we use the fact that the initial orientation of the orbit of a boulder is the same as the orientation of Dimorphos orbit. When the collision happens, a boulder is usually near a pericenter of its orbit so its velocity vector is near perpendicular to their position vector (w.r.t. Didymos), so it should have similar direction to the Dimorphos velocity vector. The faster object, in this case a boulder, must then hit the slower one from behind i.e. from the direction opposite to velocity vector of the slower object. Since the boulders are not exactly in their pericenter and their orbital orientation evolves over time, the impact locations are heavily scattered with maximum number of re-impacts around the 
($+90^{\circ}$,$0^{\circ}$) point.

The distribution of re-impacting bodies may be useful in the analysis of Hera images since some of the observed craters will date back to the time of DART impact. These fresh craters may mask some of the older features on the surface of both asteroids and, in the case of crater counting for age determination, they should be excluded from the crater distribution. Therefore, our maps of re--impactors can be a useful tool for interpreting the crater distribution of the asteroid surfaces. 

\subsection{Quasi-stable orbits}
\label{sec:quasi}

The previous dynamical analyses and simulations of the particle orbits \citep{Rossi} show that stable orbits can exist in the Didymos-Dimorphos system. The stable orbits could be either circumbinary -- orbiting both asteroids outside of the orbit of Dimorphos, or  circumprimary -- orbiting around Didymos inside the orbit of Dimorphos. Those orbits exist only if the solar radiation pressure is weak enough, which means that only larger bodies with low area-to-mass ratio should remain in the system. The existence of such orbital solutions is not a sufficient condition, there also must be a mechanism placing the boulders released from Dimorphos on this kind of orbits. Therefore, it is required to perform simulations where boulders are released from the surface of Dimorphos to test whether some of these bodies are placed in these almost stable orbits. 

\begin{figure}
   \centering
   \includegraphics[width=\hsize]{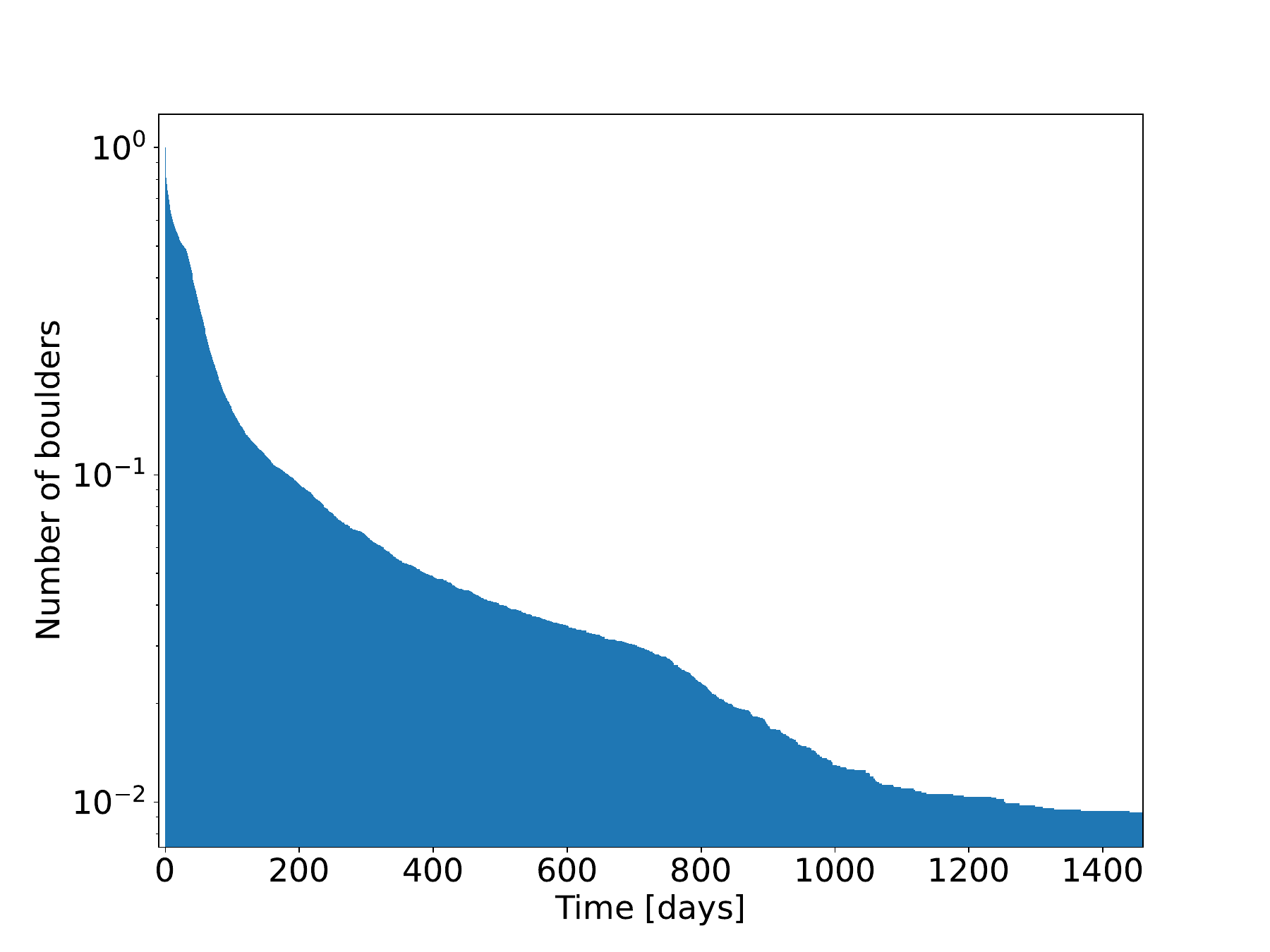}
      \caption{The total number of particles in the simulation over time (where 1 is the normalised initial population size)}. Objects that escaped or re-impacted are removed from the simulation.
         \label{fig:number_of_boulders}
\end{figure}

The large initial population of the boulders in the simulation exponentially decays over time as shown on Fig.~\ref{fig:number_of_boulders}. The rate of decay slowly decreases over time, but about 700 days after the DART impact the rate has a temporary increase. This temporary rise occurs when Didymos is near its perihelion around the Sun. After the perihelion, approximately after 1000 days from the beginning of the simulation, the decay rate slows down again. A total of 92 objects survived until the end of the simulation with 6 of them on escape trajectories that have not yet reached the limiting distance of 5 Hill radii from Didymos but are going to escape within the following 30 days. If we exclude these objects and a few others on very distant orbits which are heliocentric rather than circumbinary,  80 objects survive locked in orbit around the binary asteroid, about 0.8 percent of initial number of particles in the simulation. 

\begin{figure*}
\centering
\includegraphics[width=\textwidth]{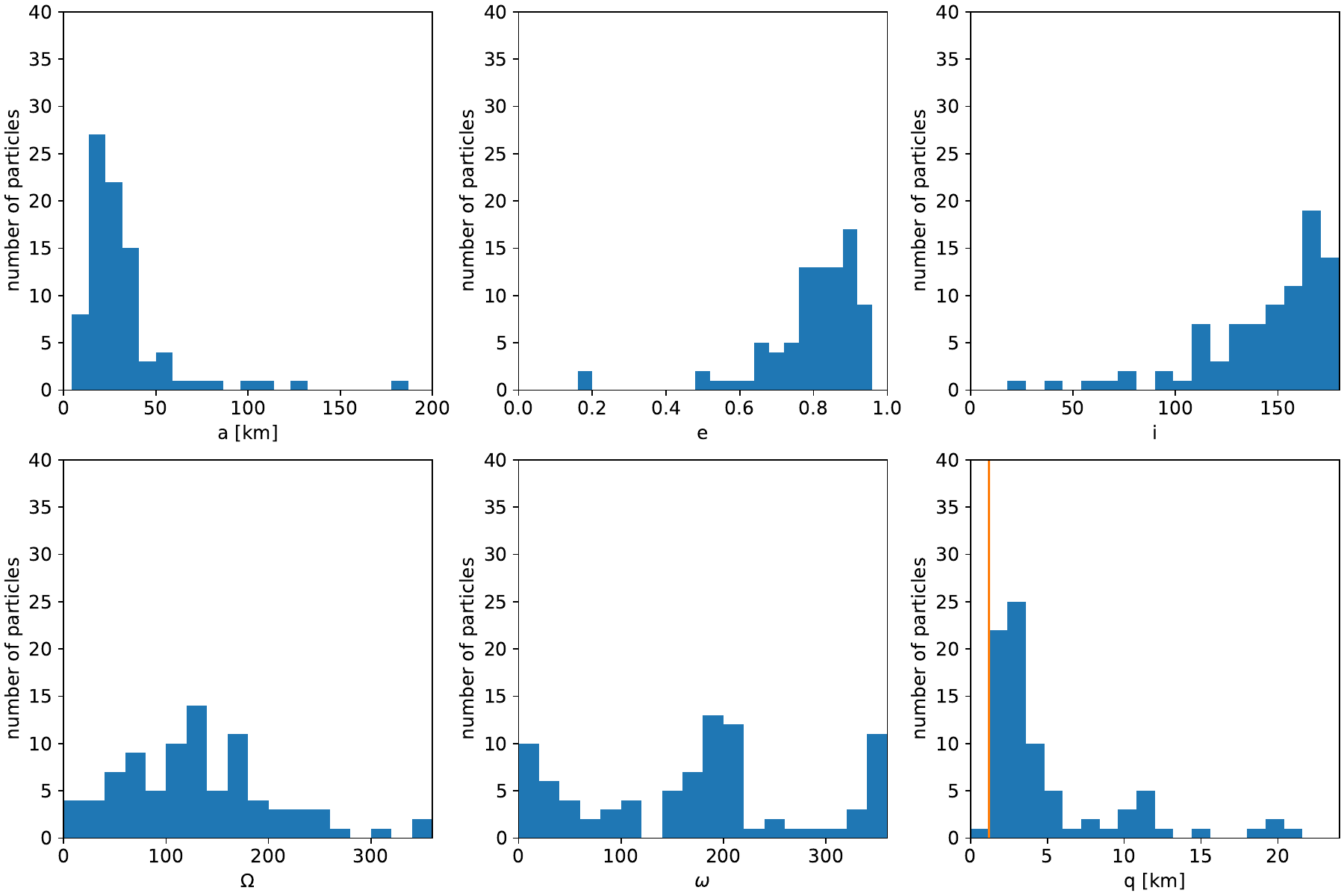}
\caption{Osculating elements distribution at the end of the simulation. Only elliptic orbits are included. In the last plot showing the distribution of pericenter distance we excluded 5 particles with $q>24$. for better readability (orange line shows the distance of 1.2 km)}
          \label{fig:elements}%
\end{figure*}

The histograms of Fig.~\ref{fig:elements} show the distribution of the osculating orbital elements of the surviving particles at the end of the simulation. The angular elements are measured in the ecliptic J2000 reference frame. It is noteworthy that all the surviving objects are on circumbinary orbits suggesting  that there are no dynamical pathways connecting the boulders from their ejection to an almost stable circumprimary orbit around Didymos lasting at least a few hundreds  days. Also there are no particles orbiting Dimorphos in the simulation. The typical orbit of long term surviving objects have high eccentricity (only 14 of the final orbits have $e<0.7$) and are on a retrograde orbits with low inclination. Prograde and inclined orbits are also present but they are less frequent. All except one of the orbits have the pericenter distance larger than the orbit of Dimorphos around Didymos. Low pericenter distances significantly reduce the survival chances due to the high probability of close encounters with the asteroids. 

The histograms of the pericenter arguments and nodal longitudes (bottom panels of Fig.~\ref{fig:elements}) show that the ascending nodes concentrates around 120 degrees while the pericenters are located close to one of the nodes i.e. $\omega$ close to  $0^o$ or $180^o$. 

%This distributions suggest that some of the particles are in a secular resonance with the Sun. However,  this guess must be taken with caution since it is based on the distribution of angular elements which are not well defined when the  inclination is close to the $180^o$. The distribution of non singular longitude of periapsis is more uniform but there is still a small peak near 120 degrees. 

\begin{figure*}
   \centering
   \includegraphics[width=0.8\textwidth]{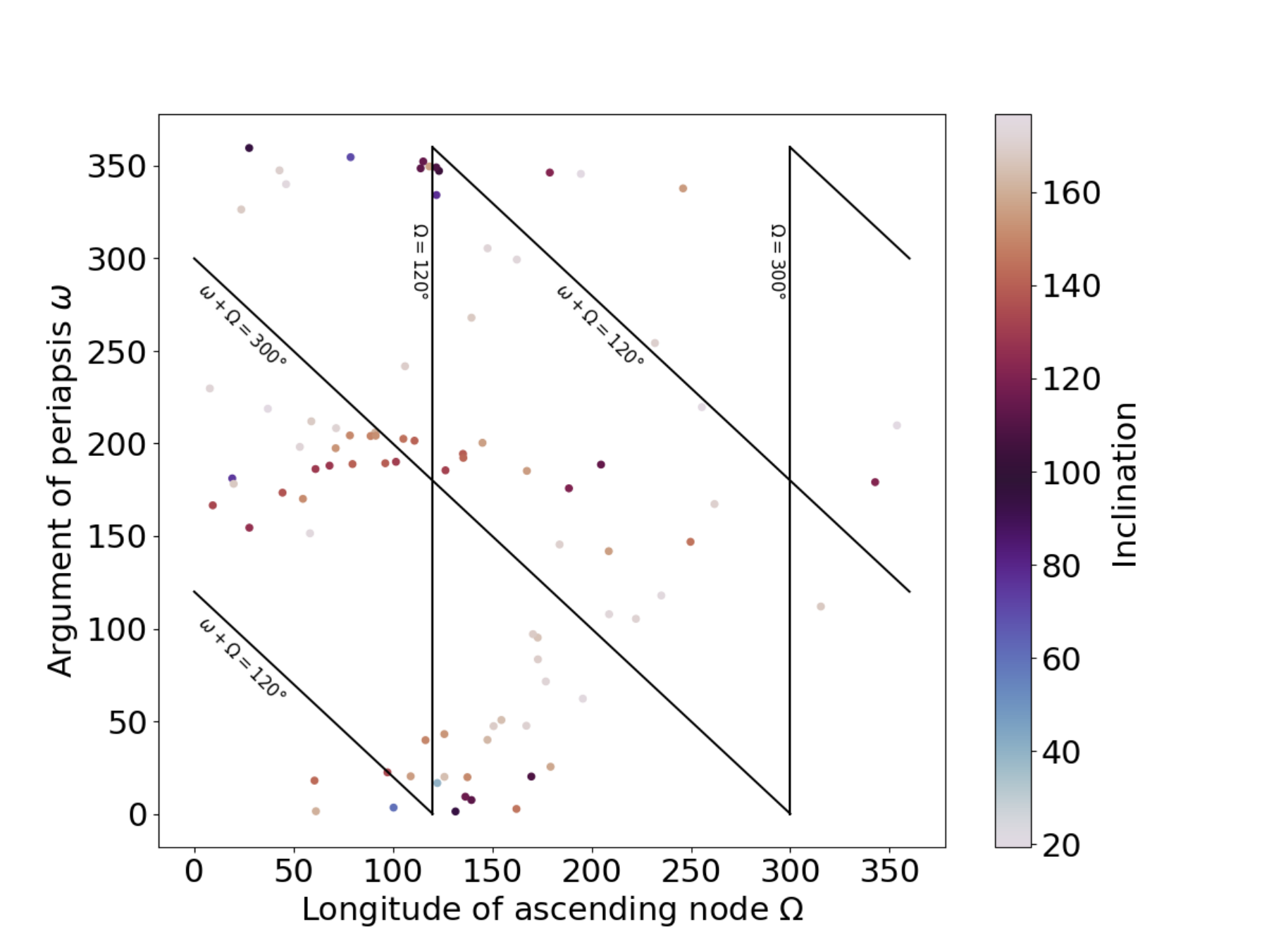}
   \includegraphics[width=0.8\textwidth]{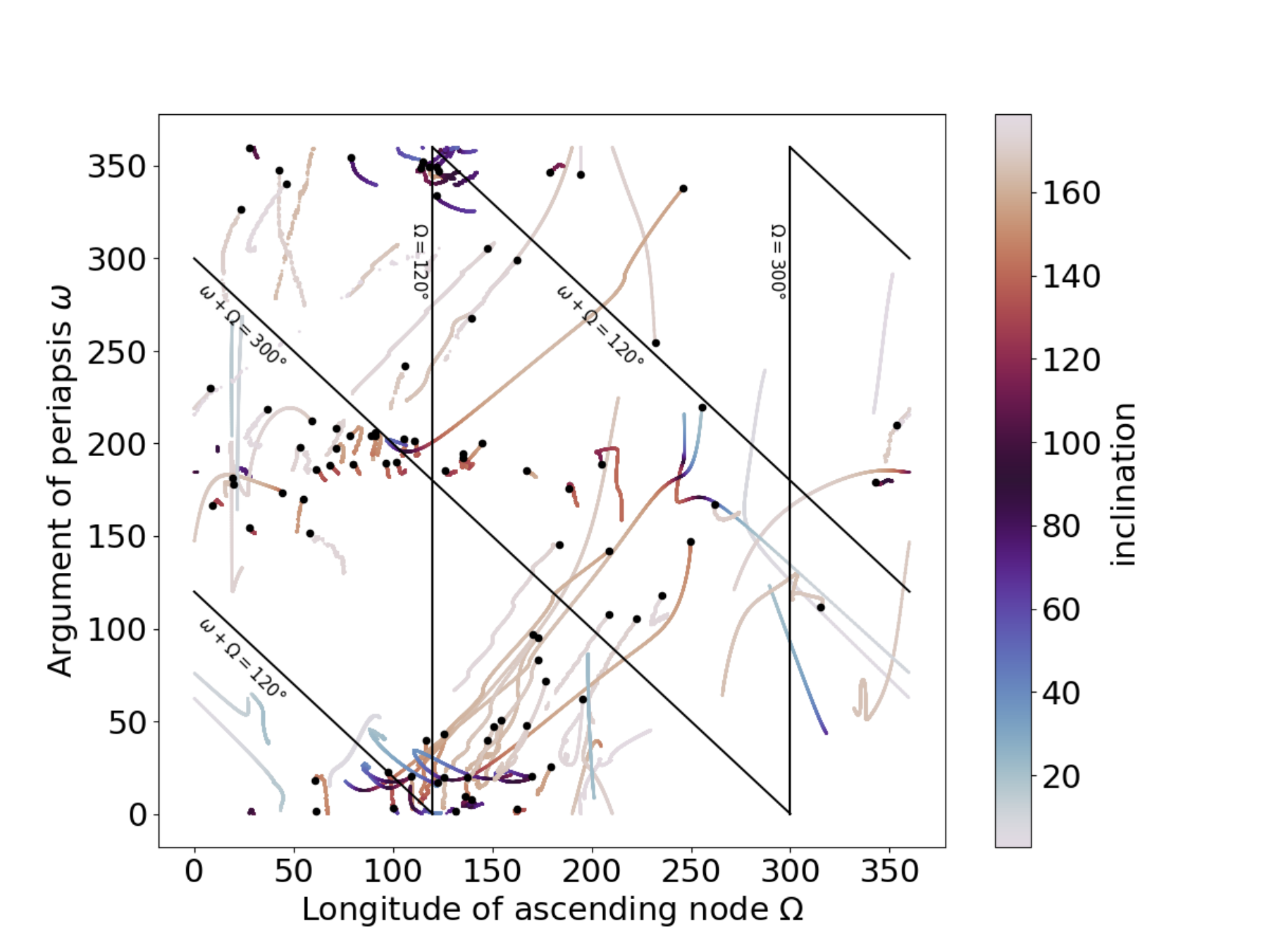}
      \caption{The final distribution (top) and evolution during last 365 days of simulation (bottom) of angular osculating elements: longitude of ascending node, argument of periapsis and inclination (as color). On the bottom figure all objects present in the last 365 days are plotted and those which are present at the end are marked by black dot at the end of their track.}
         \label{fig:omegas}
\end{figure*}

A more detailed analysis of the distribution of the orbital angles (Fig.~\ref{fig:omegas} top) show that the boulders with nodes concentrated around $120^o$ are on highly inclined orbits. Almost all bodies (with the exception of two)  with inclination close to $90^o$ have their periapsis argument concentrated around $0^{\circ}$. The evolution of the angular elements (Fig.~\ref{fig:omegas} bottom) shows that 
this might be a kind of unstable equilibrium point, where the pericenters of highly inclined orbits\footnote{The terms "high" and "low" for inclination are used to describe the angle between the orbital plane and the Ecliptic. The retrograde orbits near the Ecliptic plane with inclination close to $180^{\circ}$ are then considered to have low inclination.} are clustering and angular elements are evolving very slowly. There are also orbits moving away from this equilibrium point, lowering in the process their inclination. The direction of this equilibrium at first glance looks quite arbitrary, but it is located perpendicular to the Sun-Didymos pericenter line, which has a longitude $\approx30^{\circ}$.  The exact dynamical mechanism of this pericenter clustering requires further more detailed analysis.

\begin{figure}
   \centering
   \includegraphics[width=0.9\columnwidth]{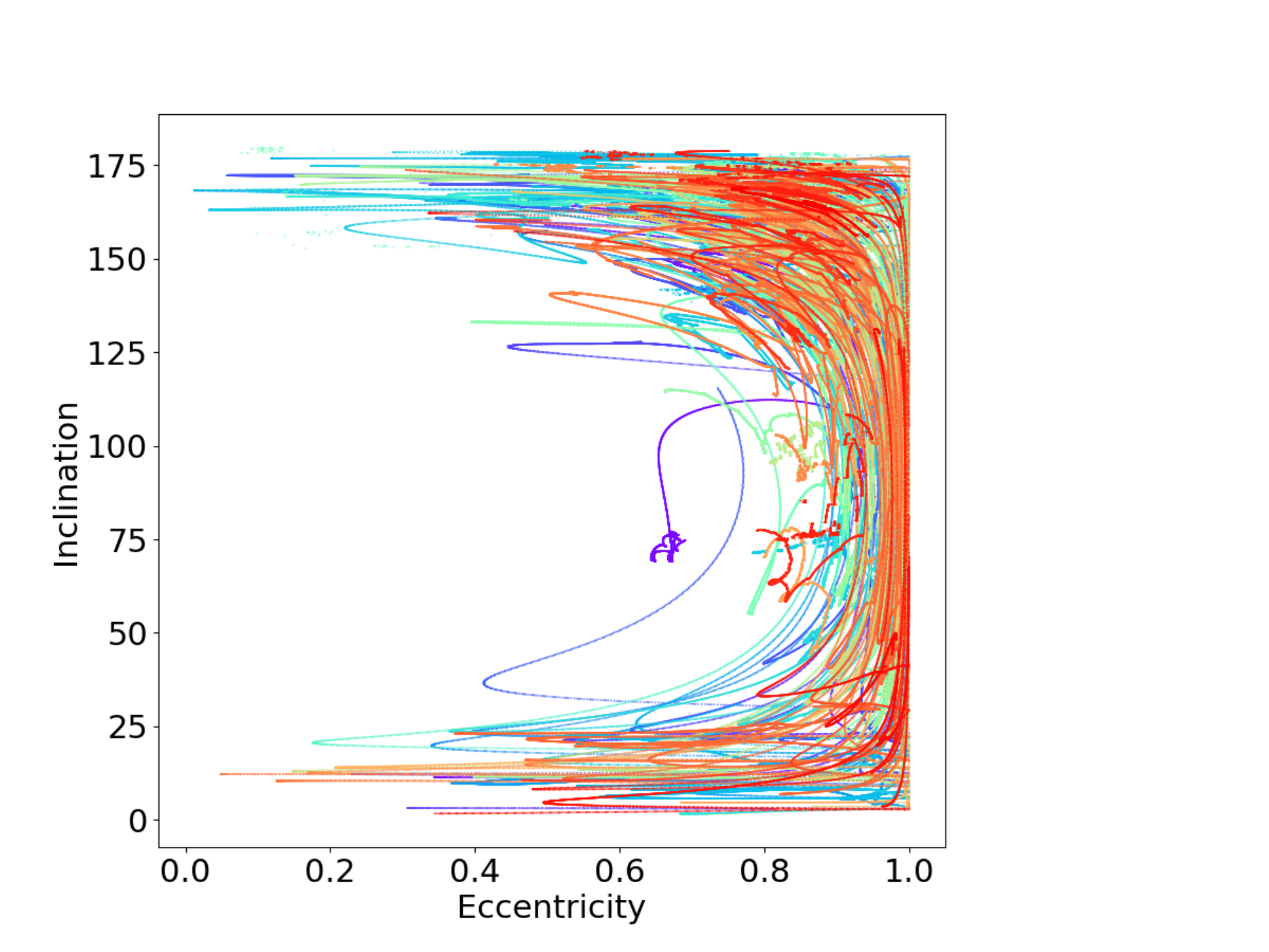}
   \includegraphics[width=0.9\columnwidth]{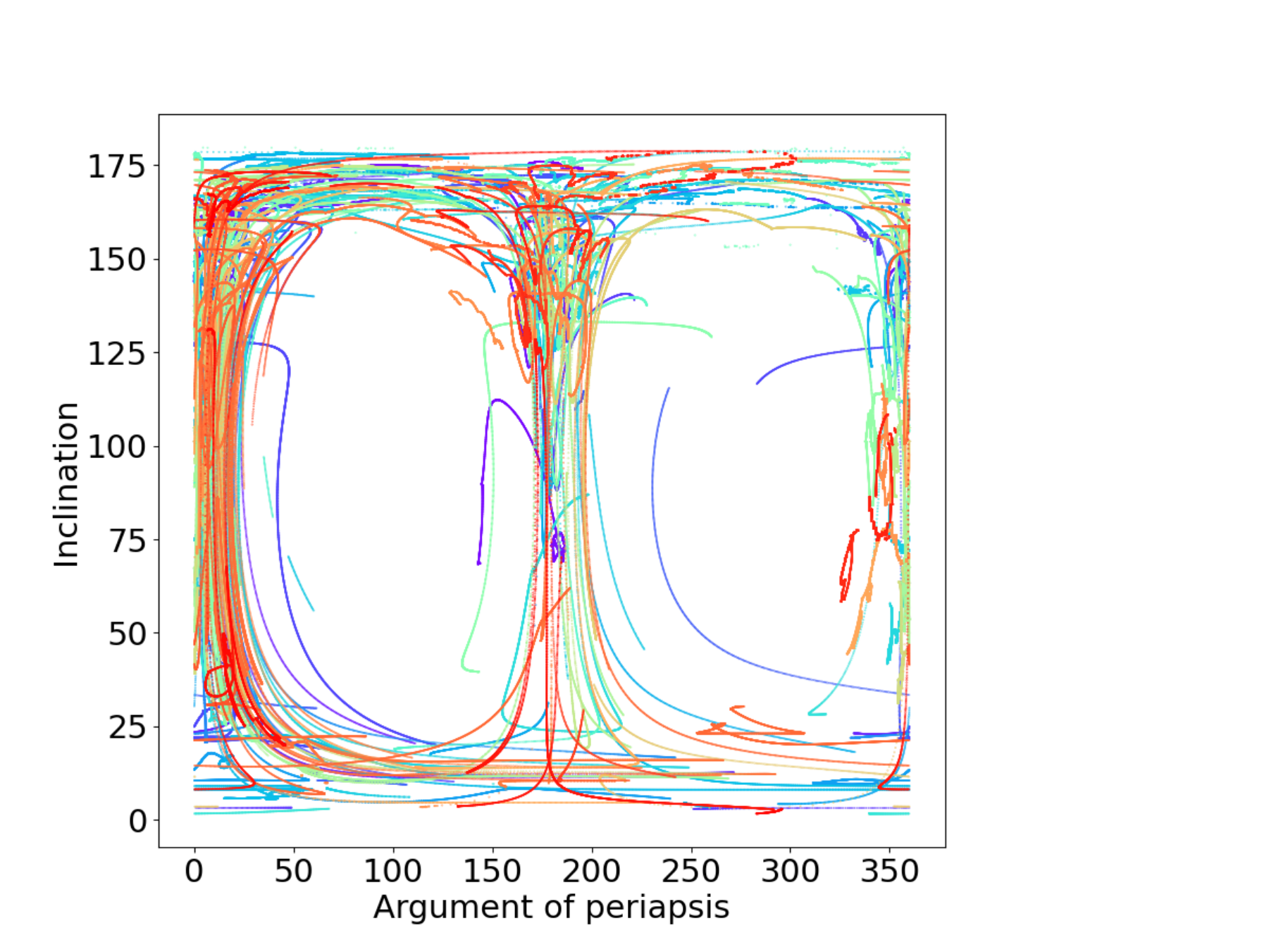}
      \caption{The trajectories on eccentricity-inclination and  inclination-argument of pericenter planes. The plots included orbits of all objects for the last two years of simulation time span. Different colours are used to distinguish different particles, but due to large number of objects the same colour may be used by more than one particle.
              }
         \label{fig:kozai}
\end{figure}

In the Fig.~\ref{fig:kozai} we show the trajectories in the inclination-eccentricity plane and in the inclination-$\omega$ plane for all particles during the last two years of the simulation. The different colours on the plots are used to distinguish between different objects. The first plot clearly shows the connection between an increase in eccentricity and inclination and also that all low eccentricity orbits are near the ecliptic plane. Both plots shows a behaviour similar to that of orbits affected by the Kozai mechanism with the Sun as the outer perturbing body. However, the orbital flips observed for many particles in the simulation are quite rapid and occur mostly near the epoch of Didymos perihelion, while in the classical Kozai mechanism the time scale should be longer than a few periods of the perturbing body. This means that the rapid orbital changes observed here are the effect of perturbations near the perihelion of the eccentric Didymos orbit. The other source of the changes in the inclination are the perturbations from Dimorphos, which may be creating an inner Kozai mechanism (\cite{naoz2017}), that affects orbits regardless of the Sun distance. The coupling of effects by inner and outer perturbing bodies together with other dynamical effects (like obliquity of the asteroids and radiation pressure) creates a complicated dynamical system, which will be explored theoretically in more detail in future works.

The vast majority of boulders have orbits entirely between the orbit of Dimorphos and the Hill radius of the Didymos system. Only one boulder at the end of simulation has pericenter closer than a Didymos orbit radius, and there are a few boulders with apocenter more distant than Hill sphere. The particles with closer apocenter have relatively strong gravitational bound to Didymos and usually are on regular elliptical orbit slowly processing due to gravitational perturbations. For those objects the more rapid changes of osculating orbits are present only as the effects of close flybys. The second group contains the high inclination orbits discussed in the previous paragraph undergoing the increase in inclination phenomena. In the final results we also find some surviving particles ejected on orbits with apocenter at very large distance (more than two times the Hill radius) from Didymos where Solar tide perturbations are considerably stronger. Those objects are loosely bound with Didymos, but for some reason they did not reach enough distance to get to our escape distance condition. Their trajectory often could have more exotic shape than weakly perturbed Keplerian ellipses and can be considered as to be on heliocentric orbit close to Didymos rather than perturbed orbit around the asteroid.  

Observing different particle trajectories in the simulation we can propose the following typical scenario for long time surviving particles: first the boulder is ejected from the surface of Dimorphos by the DART impact. If its velocity is too low it quickly re-impacts, if it is too large, the boulder escapes the system within few days or weeks. If the boulder lies in the correct velocity range it remains in the system on initial orbit with pericenter close to the DART impact location. This initial orbit can be very chaotic due to possible close encounters with Dimorphos and sooner or later the particle is either removed (escape or collision) by one of these encounters or sent into an orbit with more distant apocenter. At those larger distances Solar tide have greater effect on the orbit and could possibly increase the pericenter distance, allowing the particle to reduce the probability of close flybys near Dimorphos. This mechanism could possibly place the boulder on quasi-stable orbits where some of them can last for hundreds of days. Those quasi-stable orbits are perturbed by the Sun and asteroids gravity resulting in an increase in eccentricity and inclination or orbital flips from retrograde to prograde and vice-versa. This orbital evolution may cause the escape or re-impact of the boulder tens or hundreds of days after it was released from Dimorphos.

%ORBITAL EVOLUTION, ORBITAL FLIPS, DISTANT ORBITS...

\subsection{Size dependency}

The size of the ejecta is relevant when the solar radiation pressure becomes an important force which may significantly perturb gravity. Previous simulations \citep{Yu2017,Yu2018,Rossi} have shown that radiation pressure can quickly destabilise the orbits of small particles around Didymos--Dimorphos and then reduce their lifetime. To test the relevance of the solar radiation pressure for larger bodies we have performed an additional simulation where the size of the particles ranges from 10 cm to 10 m. In randomly creating the particle population we have used a uniform distribution of sizes to equally represent the particles in each size range. 

\begin{figure}
   \centering
   \includegraphics[width=\hsize]{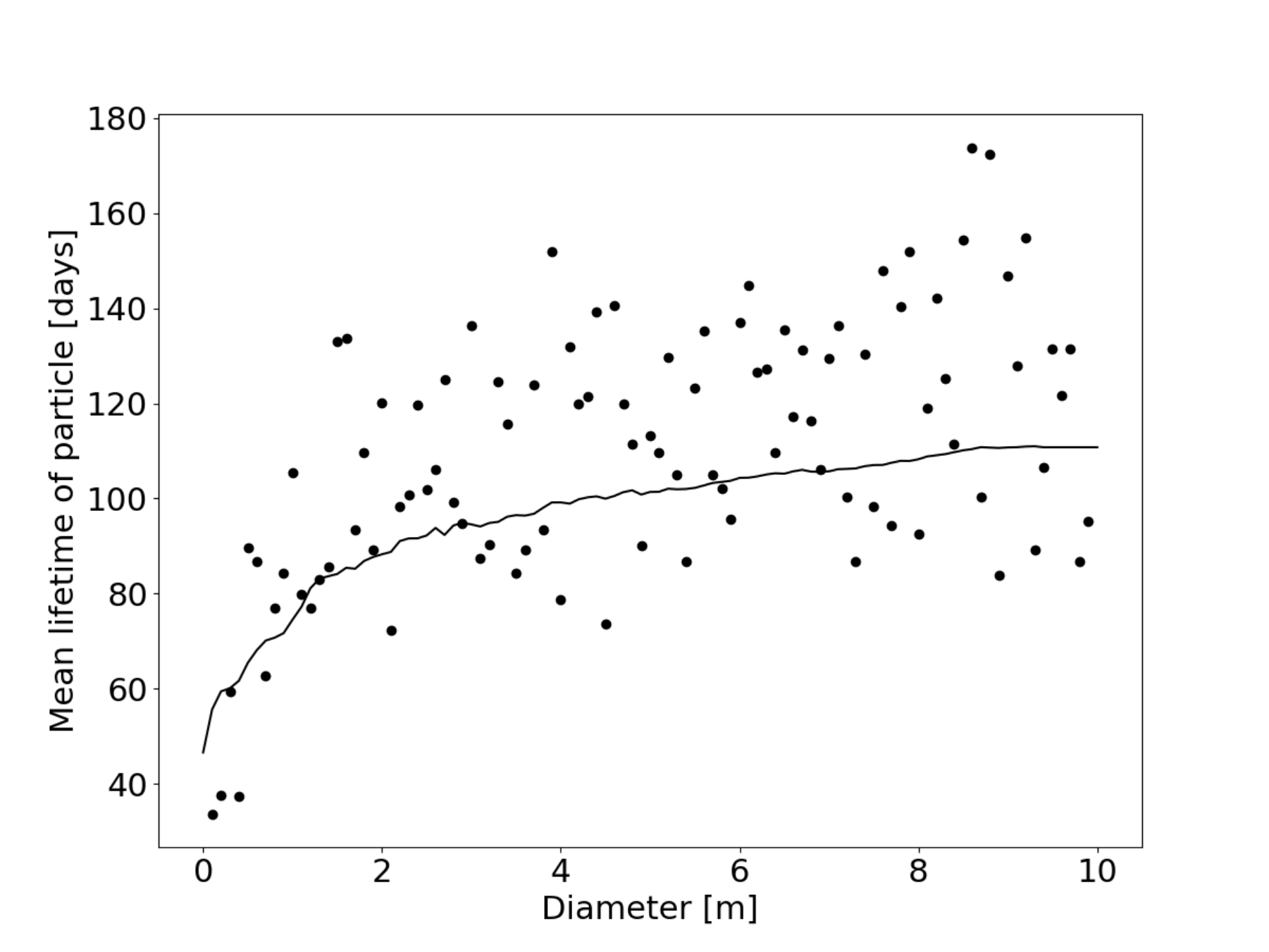}
      \caption{The mean lifetime of particles in the simulation depending on their diameter. The dots are the mean value for all the particles with size range $\pm$5 cm (about 100 particles per dot) and the solid line is the mean value with moving average for particles in a wider range of sizes $\pm$50 cm.}
         \label{fig:sizes2}
\end{figure}

In Fig.~\ref{fig:sizes2} we plot the mean particle lifetime in the simulation vs the particle size. Each point is the mean value of the lifetime for all particles encompassed within a size range of $\pm 5$ cm. The moving average with range $\pm 50$ cm (solid line in the plot) have a shape similar to a logarithmic function with a rapid initial growth for lower diameters and slower asymptotic growth for larger boulders. In the simulation the smallest particles have a lifetime less than 40 days and it raising to about 100 days for particles a few metre in size.
It must be noted that the average may depend on the initial velocities and on the condition for a particle to be considered escaped from the system. For example, particles with low initial velocity which are re-impacting Dimorphos a few hours after the DART impact can decrease the mean value. In addition, in our simulation the particles require about one month to reach the escape distance of $5 r_H$, increasing in this case the lifetime. However, the analysis of the data shows that these effects affect 'democratically' the estimates of the average lifetime.  

\begin{figure}
   \centering
   \includegraphics[width=\hsize]{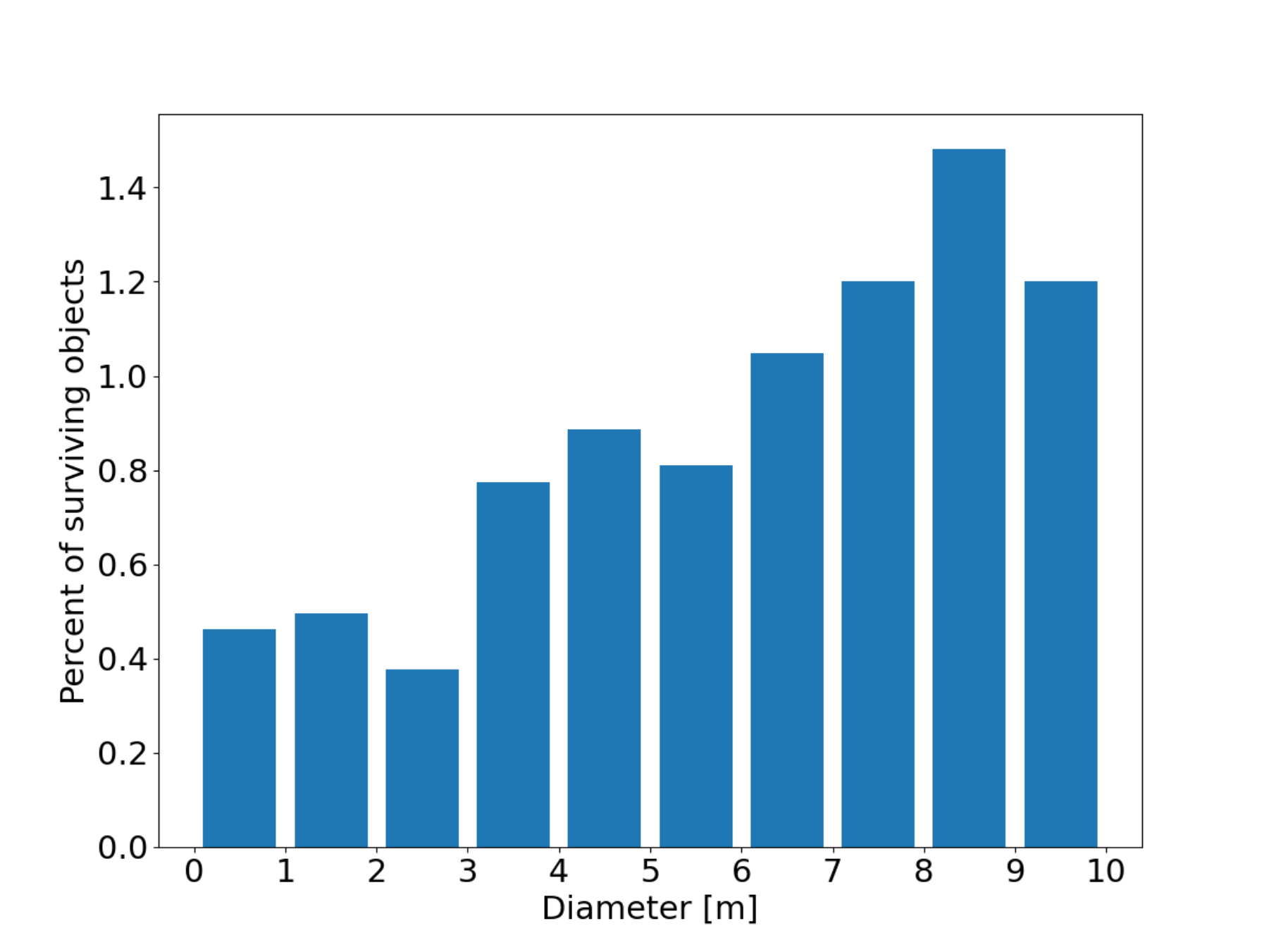}
      \caption{The percentage of boulders that survived in the Didymos System until the end of the simulation (4 years after DART impact) depending on size range.}
         \label{fig:size1}
\end{figure}

The number of objects lasting until the end of the simulation (4 Earth years) also shows a dependence on the particle size (see Fig.~\ref{fig:size1}). For particles with diameter around 4 metres the fraction of surviving boulders is close to 0.8\% in agreement with our previous computations. For larger boulders the percentage grows up to 1.2--1.4 \% while for smaller boulders it reduces to approximately 0.4 \% showing that the solar radiation pressure acts always to destabilise orbits. 

\subsection{Stability of the solutions}
\label{sec:stability}

%initial chaos, different initial conditions, clones simulations
%To estimate the stability of the long term orbits described in Sect.~\ref{sec:quasi} we performed another set of simulations where for each of the 92 long term surviving particles we added a number of clones with slightly altered initial conditions. In the first simulation we created 100 clones for each surviving particle altering the position for each clone randomly by a factor of $10^{-4}$ of the original value in each of the axis, so the absolute change was up to few millimetres. Then we used the cloned particles in simulations starting at the moment of the DART impact. 
To estimate the stability of the long term orbits described in Sect.~\ref{sec:quasi} we performed another set of simulations where we used the orbits of 92 long term surviving boulders to create a large population of clones with slightly altered initial conditions.

In the first simulation we generated 100 clones for each long surviving particle by randomly altering the initial position of the original particle by a factor of $10^{-4}$ of the original value in each of the Cartesian coordinates. Therefore, the absolute change was smaller than a few millimetres. These cloned particles are integrated starting at the moment of the DART impact.

The very small change in initial conditions has a significant impact on the resulting particle dynamics and their lifetime. During the first 30 days more than 3000 clones (i.e., 33 \% of the total) re-impacted with one of the asteroid. Further on, during the first 100 days 5322 objects are removed from the simulation due to re-impacting or escaping. This means that the rate of survival in the clone population is only slightly higher than in the original simulation and still most of the particles are removed. The number of particles surviving until the end of the simulation is 1\%, just slightly higher than the global statistics in the original simulation with random initial conditions. 
The reason of this small difference is that in the simulation  with the clones we only include objects with an initial velocity large enough to avoid immediate re-impact against Dimorphos.

To test if a specific set of initial conditions could be potentially more prone to stability, for each of the 92 long lasting initial boulders we counted the number of
clones which survived over a long time--scale. For about half (47) of the original 92 quasi--stable boulders we find that none of their clones survived at least for the same time--span of the original cloned boulder. 
For 25 original bodies only one clone survived, while for the remaining 67 cases more than 1 clone lasted till the end of the simulation. In a single case, 8 clones (out of 100) of the original body survived, while 6 clones survived in other two cases. This suggests that the dynamics is strongly chaotic and a small change (few millimetres as stated above) in the initial position can dramatically alter the lifetime of a particle in the system. In addition, we can state that, within the assumptions of our model and the explored simulation cases, there is not a specific extended set of initial conditions leading to quasi--stable orbits.

%Only 1 clone survived for 25 original bodies, for the others more than one clone last till the end of the simulation. In a single case, 8 clones of the original body (out of 1000) survived and 6 clones in other two cases. This strongly suggests that the dynamics is strongly chaotic and a small change in the initial position can dramatically alter the lifetime of a particle in the system. \textbf{In addition we can state that, within the assumptions of our model and the explored simulation cases, there is not a specific set of initial conditions granting a quasi--stable orbit.}
To test whether the particles, after avoiding the initial highly chaotic evolution, can be injected into stable orbits, we
 performed an additional set of simulations where the clones are created 100 days after the DART impact. For each of the 92 original quasi--stable objects we generated 100 clones by altering the coordinates of their position vector again by up to $10^{-4}$ of its nominal value (corresponding in this case to an absolute change of a few metres). Then we numerically integrated this new population of 9200 clones over a time-span of 4 years (100 days beyond the end of the original integration). Out of this initial population, 3285 particles survived the entire simulation (about 36\%), 909 impacted Didymos, 646 re-impacted Dimorphos and 4360 escaped from the system. 
A small number of original objects give rise to a line of clones quite stable, with most of them surviving until the end of the simulation. Indeed, 7 of them have all 100 clones surviving. On the other hand, about half of the original objects have less than 25\% of their clones lasting till the end of the simulation and 3 have all of their clones removed by escapes and re-impacts. 

%If we compare this result to the number of escape and re-impact events in the initial simulation after 100 days (i.e. last column of Table~\ref{tab:objects}), we can see that the percentage of surviving objects is much higher. If we do not count the surviving objects, the relative percentage values of each of the removal events agrees with the original simulation.  The escapes are 74 \%, the collisions with Didymos are 15\% and the collisions with Dimorphos are 11 \% of all the particle removal events, while in the original simulations we have 75\% of escapes, 15\% of Didymos collisions and 10\% of Dimorphos collisions.

The higher percentage of survivors (36\% vs 1\%) in this second population of clones implies that, after 100 days, the 92 original quasi--stable boulders have reached a zone of the phase space where it is more easy to survive in orbit for a long interval of time. This also points to the possible existence of wide regions in the phase space where quasi--stable orbits can be found, but they are not necessarily accessible by boulders ejected after a collision on the surface of Dimorphos. 

%The escape rate of clones for most of the objects is around 60\%. A further analysis of the trajectories agrees with the results of the original simulation, for example we observe an increased rate of escaping particles near the Didymos perihelion.

\section{Discussion}
\label{sec:Discussion}

An important question concerning the environment around the Didymos--Dimorphos system is the long term survival of boulders ejected after the DART impact. If they stay long enough on quasi--stable orbits they may be observed by the Hera mission but, at the same time, they would represent a threat for the spacecraft. To answer this question we have performed detailed numerical simulations of a large number of bodies in order to have a glimpse on the dynamics of the system. We have developed an accurate numerical model able to include all the relevant physics and explore a reasonable range of initial conditions for the ejecta.  We have also implemented in the code the correct geometry of ejection by using the SPICE kernels to have the exact orientation of the binary system in the appropriate reference frames. 

In general the dynamics near Didymos-Dimorphos can be viewed as a restricted four body problem. For particles closer to the asteroid pair the most relevant forces are the gravity of Didymos and Dimorphos while for the more distant ones the gravity of the Sun plays an important role. A significant aspect of the the combination of all the gravitational perturbations is that they force a relation between the eccentricity and the inclination of the boulders similar to a Kozai state but with a different behaviour. An increase in inclination is connected with an increase in eccentricity (however not all highly eccentric objects increase their inclination). Paradoxically, this may act as a sort of protection mechanism against instability since the high inclination reduces the probability of close encounter with Dimorphos when the orbits are eccentric. In addition, these perturbations enable orbital flips leading to polar and prograde orbits.

%According to our numerical modelling the long term dynamics of the boulders can be considered as the superposition of two restricted three body problems: the first is made of Didymos, Dimorphos and a test particle representing the boulder. The second is composed by the Sun, Didymos and the test particle. The gravitational perturbations force a relation between the eccentricity and the inclination of the boulders similar to a Kozai state where a high eccentricity corresponds to low inclination and vice-versa. This may act as a sort of protection mechanism against instability since when the orbits are eccentric and get close to the binary risking repeated close encounters, they are also on inclined orbits reducing the probability of close approaches with the binary asteroid. In addition, these perturbations enable orbital flips leading to polar and prograde orbits.  

In our study we show that a fraction of the ejected boulders can indeed survive and orbit around the binary asteroid for an extended period of time. To estimate the fraction of the initial boulders that are injected in these quasi--stable orbits we have integrated a large number of bodies in order to explore in more detail the phase space and to derive a statistical probability of survival. However, since we do not know from observations which is the real number of boulders ejected by the impact, we cannot predict even an approximate number of bodies that will orbit the binary at the time of the Hera arrival. The Hubble Space Telescope (HST) observed over 60 large boulders and several boulders were also observed by the LICIACube. Whereas the boulders observed by HST are mostly on escape trajectories, these observations allows us to assume that a significant number of boulders were ejected, possibly also with a low initial velocity. Some of them may end up on the quasi--stable orbits which we outline in our paper. The fraction of surviving objects in our simulation is small, about $1\%$ of the initial sample, but not negligible. 
Given the strong chaotic nature of the evolution it is hard to postulate if the use of more precise initial conditions or of an improved physical model (e.g., including the post-impact binary eccentricity or Dimorphos’s asynchronous rotation state) could lead to more realistic results. In any case, from our statistics, it can be safely inferred that there is a very limited risk of impacts between the arriving Hera spacecraft and a large object trapped within the system.

Since the long term orbits are not stable, there  is an almost  continuous leak of bodies from the population of boulders moving in the system. The escaping rate grows when Didymos approaches the Sun at pericenter due to the increased solar tide. These boulders might be observed from the ground in particular in 2025 when Didymos passes through the pericenter. 

An additional outcome of our modelling is the distribution of re--impacts on the surface of Dimorphos and impacts on that of Didymos which can be classified as sesquinary impacts. The knowledge of their distribution is useful to evaluate the extended emission of dust from the surface of the two bodies which refills the dust tail even at later times respect to the DART impact. It is also relevant when interpreting the images taken by Hera since these sesquinary impacts can slightly change the local features by increasing the density of boulders and by affecting in different ways the local surface composition.

\section{Acknowledgments}
\label{sec:Ack}
{This work is supported by the Italian Space Agency (ASI) within the Hera agreement (ASI-UNIBO, n. 2022-8-HH.0).
The authors wish to thank the anonymous reviewer for the useful and constructive remarks that helped to improve the paper.
}

\bibpunct{(}{)}{;}{a}{}{,}
\bibliographystyle{aa} 
\bibliography{biblio.bib} 
\end{document}